\documentclass{emulateapj}

\slugcomment{To appear in AJ., XX.}
\shorttitle{SN\,Ia Rates in Radio Loud Galaxies}
\shortauthors{Graham et al.}
\usepackage{graphicx}

\begin{document}

\title{The Type Ia Supernova Rate in Radio and Infrared Galaxies from the CFHT Supernova Legacy Survey}

\author{
M.L. Graham\altaffilmark{1},    
C.J. Pritchet\altaffilmark{1},  
M. Sullivan\altaffilmark{2},    
D.A. Howell\altaffilmark{3},    
S.D.J. Gwyn\altaffilmark{4},    
P. Astier\altaffilmark{5},      
C. Balland\altaffilmark{5},     
S. Basa\altaffilmark{6},        
R.G. Carlberg\altaffilmark{7},  
A. Conley\altaffilmark{8},      
D. Fouchez\altaffilmark{9},     
J. Guy\altaffilmark{5},         
D. Hardin\altaffilmark{5},      
I.M. Hook\altaffilmark{2,10},   
R. Pain\altaffilmark{5},        
K. Perrett\altaffilmark{7},     
N. Regnault\altaffilmark{5},    
J. Rich\altaffilmark{11},       
D. Balam\altaffilmark{1},                   
S. Fabbro\altaffilmark{12,1},                 
E.Y. Hsiao\altaffilmark{1,13},                 
A. Mourao\altaffilmark{12},                 
N. Palanque-Delabrouille\altaffilmark{11},  
S. Perlmutter\altaffilmark{13},             
V. Ruhlman-Kleider\altaffilmark{11},        
N. Suzuki\altaffilmark{13},                 
H.K. Fakhouri\altaffilmark{13},                  
E.S. Walker\altaffilmark{2}                  
} 

\altaffiltext{1}{Department of Physics and Astronomy, University of Victoria, PO Box 3055 STN CSC, Victoria BC V8T 1M8, Canada}
\altaffiltext{2}{University of Oxford Astrophysics, Denys Wilkinson Building, Keble Road, Oxford, OX1 3RH, UK}
\altaffiltext{3}{University of California at Santa Barbara, Santa Barbara CA 93106-9530 USA}
\altaffiltext{4}{Canadian Astronomy Data Centre, NRC Herzberg Institute for Astrophysics, 5071 West Saanich Road, Victoria BC V9E 2E7, Canada}
\altaffiltext{5}{LPNHE, CNRS-IN2P3 and Universit\'{e}s Paris VI \& VII, 4 place Jussieu, 75252 Paris Cedex 05, France}
\altaffiltext{6}{LAM, Pole de l'Etoile Site de Chateau-Gombert, 38 rue Frederic Joliot-Curie, 13388 Marseille Cedex 13, France}
\altaffiltext{7}{Department of Astronomy and Astrophysics, University of Toronto, 60 St. George Street, Toronto ON M5S 3H8, Canada}
\altaffiltext{8}{Department of Astrophysical and Planetary Sciences, University of Colorado, Boulder, CO 80309-0391}
\altaffiltext{9}{CPPM, CNRS-IN2P3 and Universit\'{e} Aix-Marseille II, Case 907, 13288 Marseille Cedex 9, France}
\altaffiltext{10}{INAF, Osservatorio Astronomico di Roma, via Frascati 33, 00040 Monteporzio (RM), Italy}
\altaffiltext{11}{CEA/Saclay, DSM/Irfu/Spp, 91191 Gif-sur-Yvette Cedex, France}
\altaffiltext{12}{CENTRA-Centro M. de Astrofisica and Department of Physics, IST, Lisbon, Portugal}
\altaffiltext{13}{Lawrence Berkeley National Laboratory, 1 Cyclotron Road, Berkeley CA 94720 USA}

\begin{abstract}

We have combined the large SN\,Ia database of the Canada-France-Hawaii Telescope
Supernova Legacy Survey and catalogs of galaxies with photometric redshifts,
VLA 1.4 GHz radio sources, and Spitzer infrared sources. We present eight SNe\,Ia
in early-type host galaxies which have counterparts in the radio and infrared source catalogs.
We find the SN\,Ia rate in subsets of radio and infrared early-type galaxies is $\sim$1--5 times
the rate in all early-type galaxies, and that any enhancement is always $\lesssim$ $2\sigma$.
Rates in these subsets are consistent with predictions of the two component ``A+B'' SN\,Ia rate model.
Since infrared properties of radio SN\,Ia hosts indicate dust obscured star formation,
we incorporate infrared star formation rates into the ``A+B'' model.
We also show the properties of SNe\,Ia in radio and infrared galaxies
suggest the hosts contain dust and support a continuum of delay time distributions
for SNe\,Ia, although other delay time distributions cannot be ruled out based on our data.

\end{abstract}


  \section{INTRODUCTION}
  \label{S1Intro}
  
  Type Ia supernovae (SNe\,Ia) are currently the best cosmological tool to investigate the
  dark energy driving the accelerated expansion of the universe \citep{R98,P99,A06,W-V07,R07}.
  Though generally accepted as thermonuclear explosions of carbon-oxygen white dwarfs which
  have accreted sufficient mass from their companion, two competitive models exist for the
  companion scenario: the single-degenerate (main sequence or red giant companion) and
  double-degenerate (another white dwarf). Since the time delay between star formation and
  SNe\,Ia explosion is scenario-dependent, rates in hosts of varying star formation rate (SFR)
  help constrain this. In fact, SN\,Ia rates are known to be correlated with host morphology,
  B-K color, and by inference also with star formation \citep{Mann05}. As such, a galaxy's
  SN\,Ia rate is commonly expressed as the sum of a ``delayed'' component from old stellar
  populations and a ``prompt'' component from young stellar populations. These components are
  parametrized as ``A'' and ``B'', proportional to a galaxy's mass and SFR respectively \citep{SB05,S06a}.
  This ``A+B'' two-component model can be matched with progenitor populations that have
  distributions of delay times as described by Mannucci et al. (2006), and Pritchet et al. (2008).
  
  The C99 catalog matched with Very Large Array (VLA) 1.4GHz
  radio data revealed the specific SN\,Ia rate may be enhanced by 2--7 times 
  in radio-loud ($\rm L_{1.4\,GHz} > 10^{29}\ ergs\ s^{-1}\ Hz^{-1}$) over
  radio-quiet ($\rm L_{1.4\,GHz} < 4\times10^{27}\ ergs\ s^{-1}\ Hz^{-1}$)
  early-type galaxies (Della Valle et al. 2005, hereafter DV05).
  Although winds from active galactic nuclei (AGN) could increase accretion
  rates of the interstellar medium (ISM) onto white dwarfs, a process thought
  to trigger classical nova eruptions \citep{L02,M07}, DV05 reject this explanation
  for the enhanced specific SN\,Ia rate in radio-loud galaxies.
  They find galaxy interactions are most likely to cause the radio emission
  and supply the necessary SN\,Ia progenitors via stellar capture during
  dwarf accretion or star formation induced by major mergers \citep{DV05}.
  
  The existence of bright infrared counterparts for radio galaxies has been
  well documented and often attributed to dust obscured star formation
  coeval with the AGN  \citep{Mag08,Main08}.
  By matching a hypothetical galaxy interaction and AGN activity timeline
  with a recurring star formation model of ten $\rm 10^8$ year long episodes each separated
  by $\rm 10^{9}$ years, Mannucci et al. (2006) find the enhanced specific SN\,Ia rate in
  radio-loud early-type galaxies is best fit by a bimodal delay-time distribution (DTD)
  in which the ``prompt'' (B) delay time is constrained to just $\rm \lesssim10^8$ years.
  They suggest this implies two physical populations of SN\,Ia progenitors, but note a
  broad single-population DTD could not be ruled out. As this controversial implication
  relies on a rate enhancement found with 21 photometrically identified SNe\,Ia from
  DV05, we look to confirm this in the large database of spectroscopically typed SNe\,Ia
  from the Canada-France Hawaii Telescope (CFHT) Supernova Legacy Survey (SNLS).
  
  We use photometric redshift galaxy catalogs \citep{I06} and 1.4 GHz radio catalogs
  to calculate the SN\,Ia rate in radio-loud early-type galaxies, and also infrared
  source catalogs to look for obscured star formation in the radio-loud SN\,Ia host galaxies.
  Dust extinction in starbursts hinders SN detection: only one SN\,Ia has
  been detected, and the SN\,Ia rate remains unconstrained in starburst
  galaxies \citep{Mann03,Mann07}. We use the infrared catalogs
  to calculate the SN\,Ia rate in bright and luminous infrared galaxies,
  which are known to experience bursts of star formation
  up to $\rm 100-1000\ M_{\odot}\ y^{-1}$.
  We also statistically compare the SN\,Ia rates in radio and infrared
  galaxies to predictions of the two-component ``A+B'' model, and
  perform an identical analysis with the low redshift C99 catalog.
  
  In \S~\ref{S2Obs} we describe the SNLS and C99 SN\,Ia databases and
  galaxy catalogs, the overlapping radio and infrared source catalogs,
  and present the radio and infrared properties of SNLS and C99 SN\,Ia early-type host galaxies.
  We derive the SN\,Ia rate in radio and infrared galaxies in \S~\ref{Srates},
  compare to expectations of the two-component ``A+B'' model in \S~\ref{SABP},
  and reanalyze with relaxed data constraints in \S~\ref{Srelax}.
  In \S~\ref{SSNProp} we juxtapose the properties of the SNe\,Ia in radio and infrared
  host galaxies with known correlations between SN\,Ia properties and stellar populations.
  Finally, in \S~\ref{SDisc} we discuss the implications of our results regarding current
  SN\,Ia science and future surveys, and conclude in \S~\ref{SConc}.

  \section{OBSERVATIONS}
  \label{S2Obs}
  
  Here we describe the combination and processing of
  private and public catalogs of supernovae,
  galaxies, and radio and infrared sources.

  \subsection{CFHT Supernova Legacy Survey Catalog}
  \label{sS2SNLS}
 
  From 2003 to 2008, CFHTLS Deep component imaged four 1 deg$^2$ fields (D1--D4) every
  three to four nights (during dark/grey time, when visible) in four MegaCam filters
  ($g_M,r_M,i_M,z_M$) to a depth $i_M \simeq 25$; this was accompanied by a strong
  spectroscopic campaign to follow up on as many potential SN\,Ia candidates as possible \citep{P09}.
  After final reductions of all data, the SNLS\footnote{http://cfht.hawaii.edu/SNLS}
  will have discovered and identified hundreds of SNe\,Ia, provide the best
  direct constraints on the dark energy equation of state parameter $w$ \citep{A06}
  and be a useful compliment to other cosmological techniques \citep{K09}.
  We use the 91 and 64 SNe\,Ia identified prior to 2006 Dec 31 in
  CFHTLS deep fields 1 and 2 respectively ($\sim 30\%$ of the final four-field total),
  because suitable VLA radio surveys cover D1 and D2 only, and up to this date
  the SNLS spectroscopic analysis and typing is complete.

  \subsection{Ilbert et al. (2006) Galaxy Catalog}
  \label{sS2IlbCat}
  
  The Ilbert et al. (2006) galaxy catalog\footnote{http://terapix.iap.fr}
  incorporates VIMOS VLT Deep Survey spectroscopic redshifts to calibrate their spectral energy
  distribution (SED) fitting routine, resulting in
  accurate ($\rm \sigma_{\Delta z / (1+z)} =0.029$) photometric redshifts for
  galaxies in the four SNLS Deep fields. In optimizing the photo-z calculation,
  the accuracies of the SEDs are compromised (Ilbert, private communication) and the distribution of 
  SED types is discontinuous.  To solve this we fit 51 SEDs,
  interpolated from Coleman et al. (1980) and Kinney et al. (1996)
  templates made by Stephen Gwyn at the Canadian Astronomical
  Data Centre\footnote{http://www.cadc.hia.nrc.gc.ca/community/CFHTLS--SG/docs/cfhtls.html},
  to the catalog galaxies, and estimate galaxy stellar masses and star formation rates
  using fits of this library of SEDs to the models of Buzzoni (2005). We correct these for
  systematic offsets (of about a factor of ~2) to agree with the PEGASE models \citep{S06photoid}.
  Our SED types include E/S0 (ellipticals or early-types), Sbc and Scd (spiral galaxies),
  Irr (irregular galaxies), and SB (starbursts, $\rm sSFR\gtrsim30\times10^{-10}\ y^{-1} $).
  To ensure catalog purity we restrict the
  catalog as recommended by Ilbert et al. (2006), and limit to $i_M < 25$.

  \subsection {Additional Galaxy and SN Catalog Processing}
  \label{sS2add} 

  We identify SN\,Ia hosts as the nearest Ilbert catalog galaxy unless there
  are two within 2 arcseconds, in which case redshift is used to discriminate.
  In D1 and D2, 22 and 15 SNe\,Ia have no catalog
  galaxy within 5 arcseconds (maximum host offset) and cannot be used;
  this mainly includes SNe\,Ia in hosts which do not meet our catalog
  restrictions (17 and 12), but also a few SN\,Ia whose hosts are not in
  the original galaxy catalgs (5 and 3; many are near field edges, or
  close to foreground stars).
  For SNe\,Ia with Ilbert catalog hosts, iterative outlier rejection 
  is applied to the residual dispersion between host photometric and
  SN\,Ia spectroscopic redshifts for 
  each deep field, resulting in photometric redshift uncertainties of 
  $\rm \sigma_{1}=0.032$ and $\rm \sigma_{2}=0.028$,
  rejecting 2 and 6 SNe\,Ia hosts as outliers in D1 and D2.
  We use SN\,Ia with spectroscopic redshifts $\rm z \lesssim 0.6$, the limit 
  to which the SNLS SN\,Ia sample is nearly complete \citep{N06}.
  These restrictions yield 33 and 29 usable SNe\,Ia in D1 and D2 respectively;
  possible effects of the redshift restriction are considered in \S~\ref{Srelax}.

  For every galaxy we calculate its SN\,Ia observed rate per year, $\rm R_{Ia}$, from the two-component ``A+B'' model
  applying the time dilation correction to our observed frame of reference, as shown in Equation \ref{EAB}:
  
  \begin{equation}
    \label{EAB}
    R_{Ia} = \frac{A \times M \ + \ B \times SFR}{1+z},
  \end{equation}

\noindent
  where $\rm M$ is stellar mass ($\rm M_{\odot}$), $\rm SFR$ is star formation rate
  ($\rm M_{\odot}$ / yr), and the A and B values are from Sullivan et al. (2006):
  $\rm A = 5.3 \pm 1.1 \times 10^{-14} \ SNe \ y^{-1} \ M_{\odot}^{-1}$
  and $\rm B = 3.9 \pm 0.7 \times 10^{-4} \ SNe \ y^{-1} \ (M_{\odot} \ y^{-1})^{-1}$.
  The rate $\rm R_{Ia}$ is the number of SNe\,Ia expected in a galaxy per year of observing. We
  use it to determine an effective control time, $\rm C$, to account for SNLS survey
  efficiencies such as detection and spectroscopic completeness, and the lengths of SNLS observing seasons
  (approximately 6 months), for fields D1 and D2 separately (i.e. to have $\rm C_{1}$ and $\rm C_{2}$) as follows:

  \begin{equation}
    \label{EC}
    C = \frac{N_{obs}}{\sum{R_{Ia}}},
  \end{equation}
 
\noindent
  where $\rm N_{obs}$ is the total number of
  SNLS SNe\,Ia observed in the field prior to 2006 Dec 31 with $\rm z<0.6$, and $\rm \sum{R_{Ia}}$ is the total number
  of SNe\,Ia expected in all $\rm z<0.6$ galaxies per year of observing the field.
  We find $\rm C_1=1.580$ and $\rm C_2=1.284$ where $\rm C$ has units of years.

  \subsection{VLA Radio Catalogs}
  \label{sS2radio}

  The VLA-VIRMOS 1.4 GHz Deep survey covers D1 to a  $\rm S_{1.4GHz}$ flux of $\rm 80\ \mu Jy$, with a mean
  rms noise $\rm \sigma \simeq 17\ \mu Jy$, and 75\% completeness at fluxes $\rm S_{1.4GHz}= $80--180$\rm\ \mu Jy$ \citep{B03}.
  Of the 1054 radio sources, 503 optical counterparts are identified as the closest
  galaxy within 2 arcseconds (maximum VLA positional error). Away from the galaxy catalog's masked regions
  around foreground stars our recovery rate is $\rm \sim 60\%$, similar to Ciliegi et al. (2005).
  The VLA-COSMOS 1.4 GHz Large Project covers D2 to $\rm S_{1.4GHz} \sim45 \ \mu Jy$
  with a mean rms noise of $\rm \sigma \sim15 \ \mu Jy$ \citep{Sch06}, for which we
  match $\rm \sim 50\%$ of radio sources with a galaxy.

\begin{figure}
    \begin{center}
      \plottwo{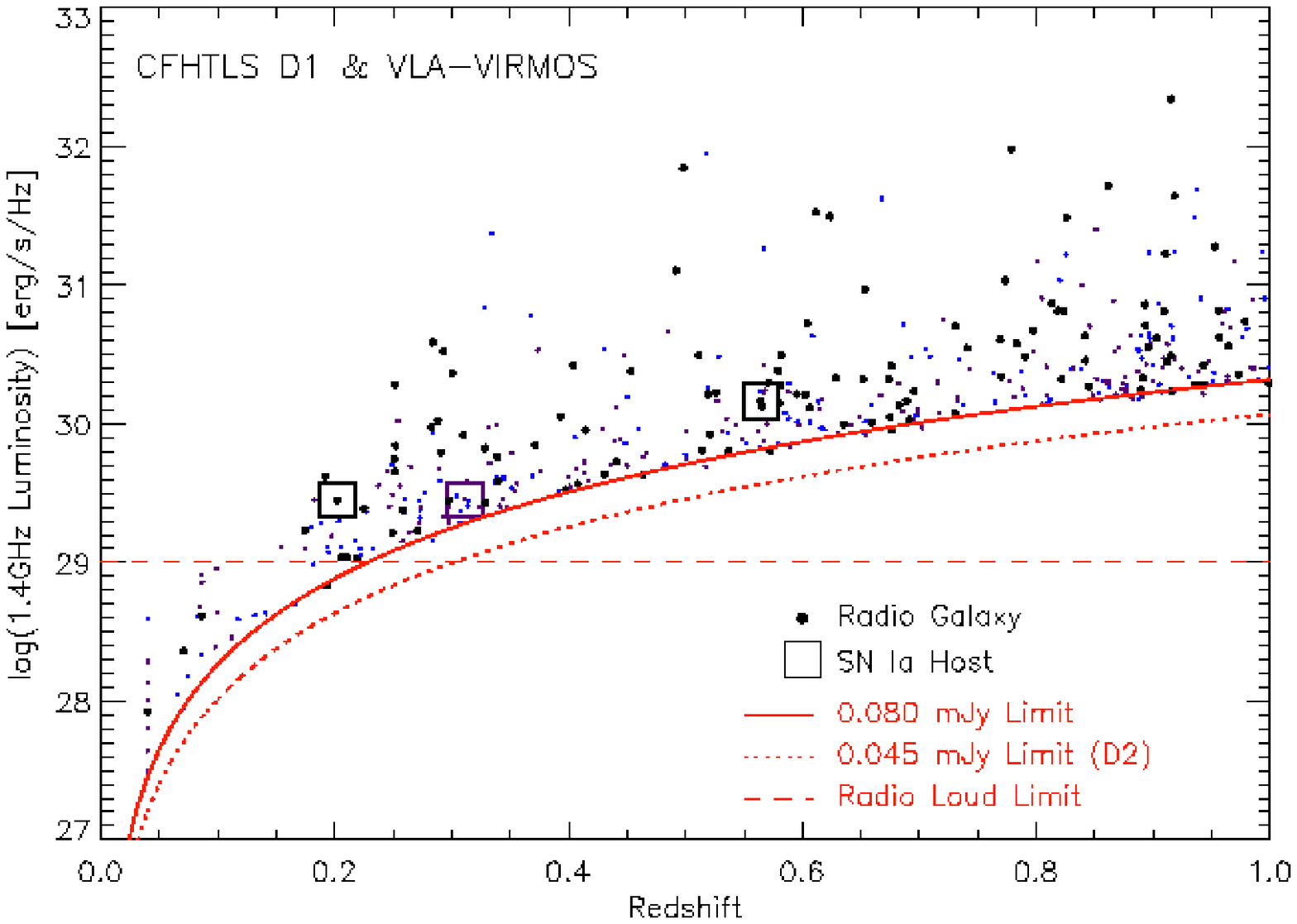}{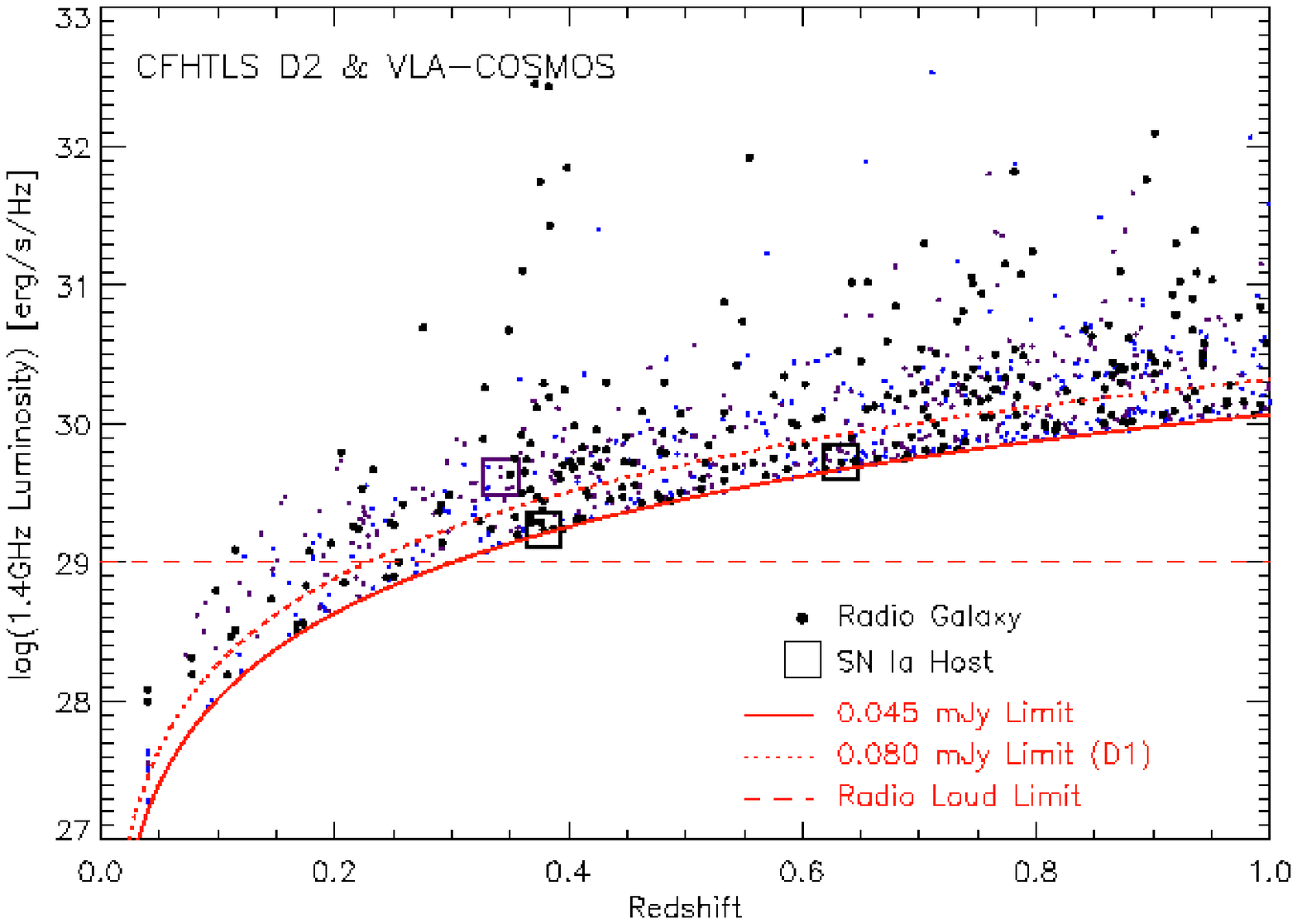}
      \caption{ $\rm L_{1.4\ GHz}$ versus photometric redshift for galaxies
        in D1 with VLA-VIRMOS (left) and D2 with VLA-COSMOS (right) counterparts.
        Filled circles are galaxies of SED type E/S0 (black), Sbc (purple), and
        later types (blue); squares are SN\,Ia hosts. Solid red line marks survey
        $\rm S_{1.4GHz}$ flux detection limit (dotted red for limit of other field), dashed red line
        for radio-loud limit. \label{FS2RlumVSz}}
    \end{center}
  \end{figure}

  Radio luminosities are derived from photometric redshifts of galaxy counterparts,
  include a $\rm (1+z)^{-1}$ bandpass correction \citep{Hogg02}, and are plotted in Figure \ref{FS2RlumVSz}.
  A galaxy is radio-loud if $\rm L_{1.4\,GHz} > 10^{29}\ ergs\ s^{-1}\ Hz^{-1}$,
  the faint-end limit of the radio luminosity function, although exact limits for radio-loud,
  -faint, and -quiet galaxies are not universal \citep{Zamfir08}.
  Ledlow \& Owen (1996) find $\rm 14\%$ ($\rm \pm2.4\%$) of elliptical galaxies with
  absolute R magnitude $\rm M_R \lesssim -20.5$ are radio-loud, and amongst
  C99 elliptical galaxies DV05 find $\rm 12\%$ ($\rm \pm2\%$) are radio-loud. In our sample of SED type
  E/S0 galaxies with $\rm z<0.6$ we find 4--8\% of galaxies with $\rm M_V \lesssim -20$
  are radio-loud, suggesting we can identify half of the radio-loud population amongst
  optically bright galaxies due to our higher minimum radio flux limit.

  The apparent bimodality of VLA-COSMOS (D2) radio luminosities in Figure \ref{FS2RlumVSz}
  is a result of combining the integrated fluxes of resolved and unresolved sources.
  The flux of the bimodality valley corresponds to the lower limit of integrated fluxes
  for resolved sources, $\rm S_{1.4GHz} \sim 0.08\ mJy$, as shown in the right-hand plot of Figure 17
  from Schinnerer et al. (2006).
  Coincidentally, it also corresponds to the VLA-VIRMOS (D1) lower limit.
  This may indicate VLA-VIRMOS and VLA-COSMOS sample slightly different radio source populations,
  with more faint, unresolved sources in D2. We consider any influence of this on
  our results in \S~\ref{Srelax}.

  \subsection{Spitzer Infrared Catalogs}
  \label{sS2IR}

  The Spitzer Wide-area Infrared Extragalactic (SWIRE) survey covers D1
  to a flux of $\rm S_{3.6\mu m}\sim6.6\ \mu Jy $ (where subscript on S denotes wavelength),
  and to $\rm S_{24\mu m}\sim300\ \mu Jy $ \citep{Lons03}.
  SWIRE objects with $\rm S_{3.6\mu m} > 200\ \mu Jy$ and stellarity $\rm >0.9$
  are most likely stars or QSOs and rejected from the
  catalog\footnote{http://irsa.ipac.caltech.edu/data/SPITZER/SWIRE/docs/delivery\_doc\_r2\_v2.pdf, page 43}.
  The Spitzer Cosmic Evolution Survey (S-COSMOS) covers D2 to $\rm S_{3.6\mu m}\sim 0.8\ \mu Jy $
  and $\rm S_{24\mu m}\sim 300\ \mu Jy $ after required aperture corrections \citep{Sanders07}.
  Objects flagged as likely compromised by nearby bright sources are
  rejected\footnote{http://irsa.ipac.caltech.edu/data/COSMOS/gator\_docs/scosmos\_irac\_colDescriptions.html}.
  Both catalogs are available via the NASA Infrared Space Archive\footnote{http://irsa.ipac.caltech.edu}.
  We reject foreground objects and QSOs, identify optical counterparts as in \S~\ref{sS2radio},
  and limit to $\rm z \lesssim 0.6$ as in \S~\ref{sS2radio}.
  The fraction of galaxies infrared counterparts in all four IRAC bands {\it and} MIPS is
  $\rm \sim2-3\%$; these are plotted on infrared color-color diagrams in Figure \ref{FS2IRcc}
  which also show AGN boundaries derived from Spitzer First Look Survey
  data of comparable flux-density limits ($\rm S_{3.6\mu m}\sim 7\ \mu Jy$ and $\rm S_{24\mu m}\sim 300\ \mu Jy$)
  and redshifts ($\rm z \lesssim 0.7$) \citep{Lacy04}. Sajina et al. (2005)
  show the region on the IRAC color-color plot (left) is actually appropriate for AGN of redshifts $0-2$. 
  It should be noted that not all galaxies in these regions are AGN-dominated at IR wavelengths.

\begin{figure}
    \begin{center}
      \plottwo{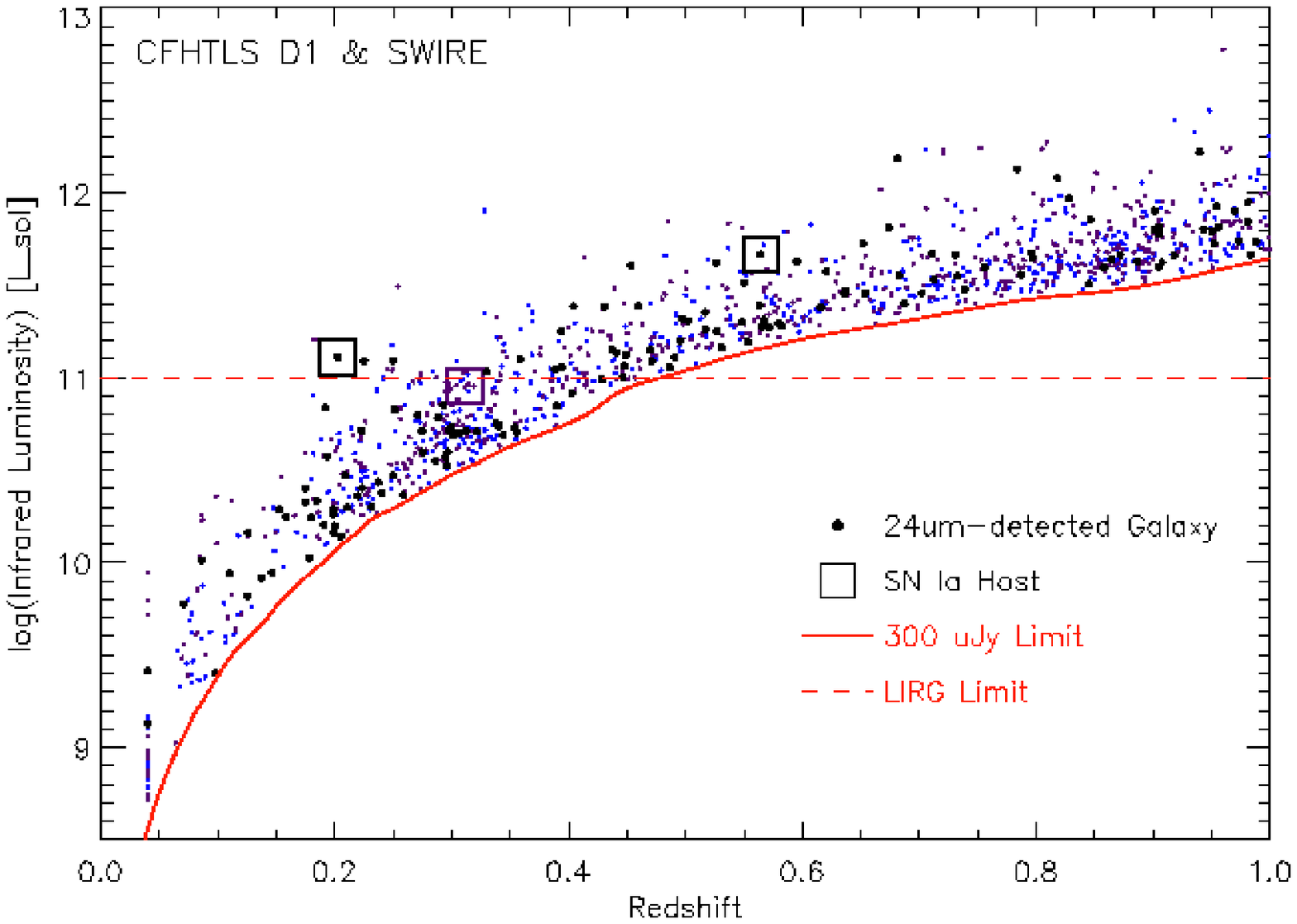}{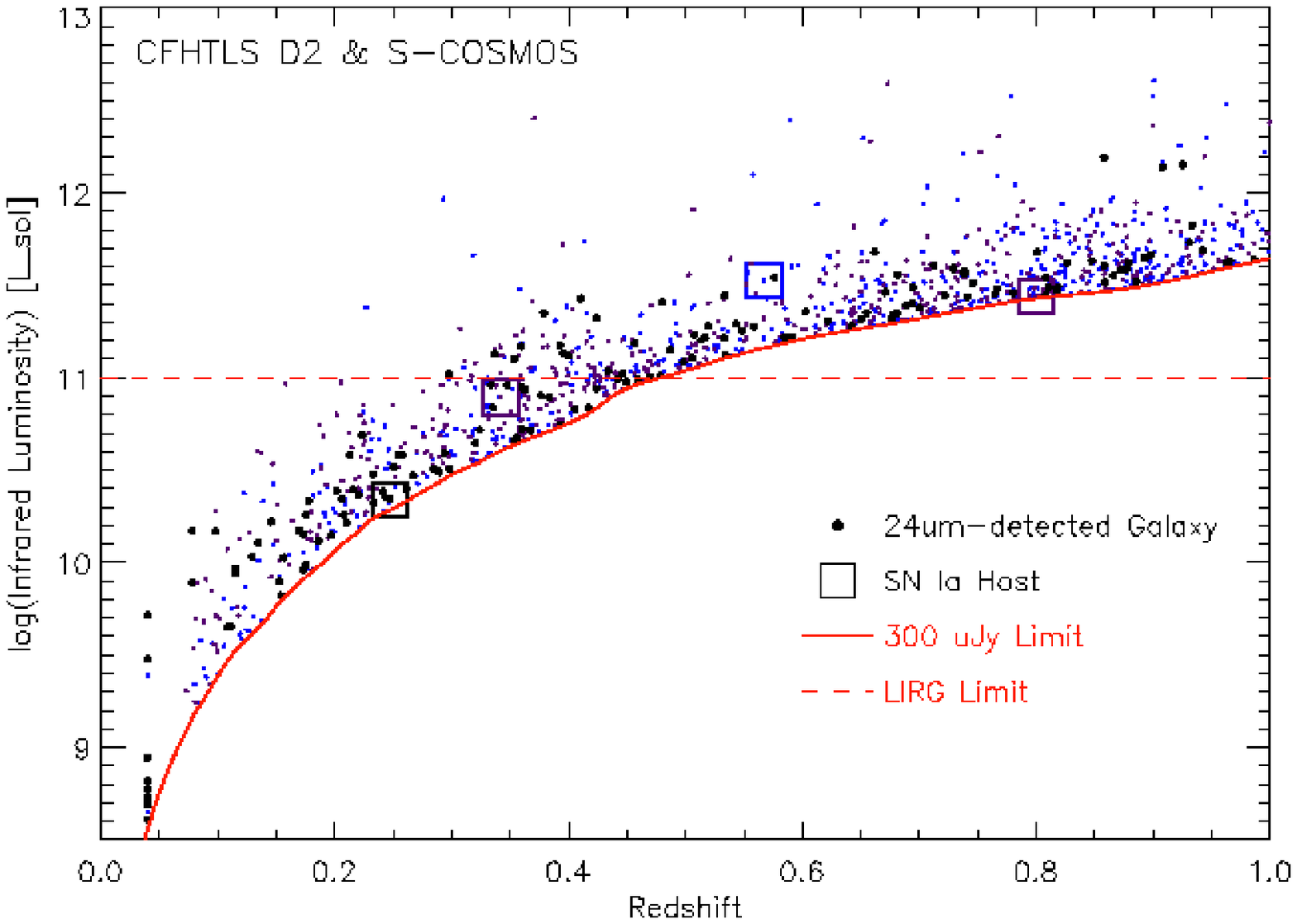}
      \caption{$\rm L_{IR}$ versus photometric redshift for Spitzer MIPS-detected galaxies
        in D1 (left) and D2 (right). Filled circles are galaxies of SED type E/S0 (black),
        Sbc (purple), and later types (blue); squares are SN\,Ia hosts. Solid red line
        marks survey $\rm S_{24\mu m}$ flux detection limit, dashed red line for LIRG limit. \label{FS2IRlumVSz}}
    \end{center}
  \end{figure}

\begin{figure}
    \begin{center}
      \plottwo{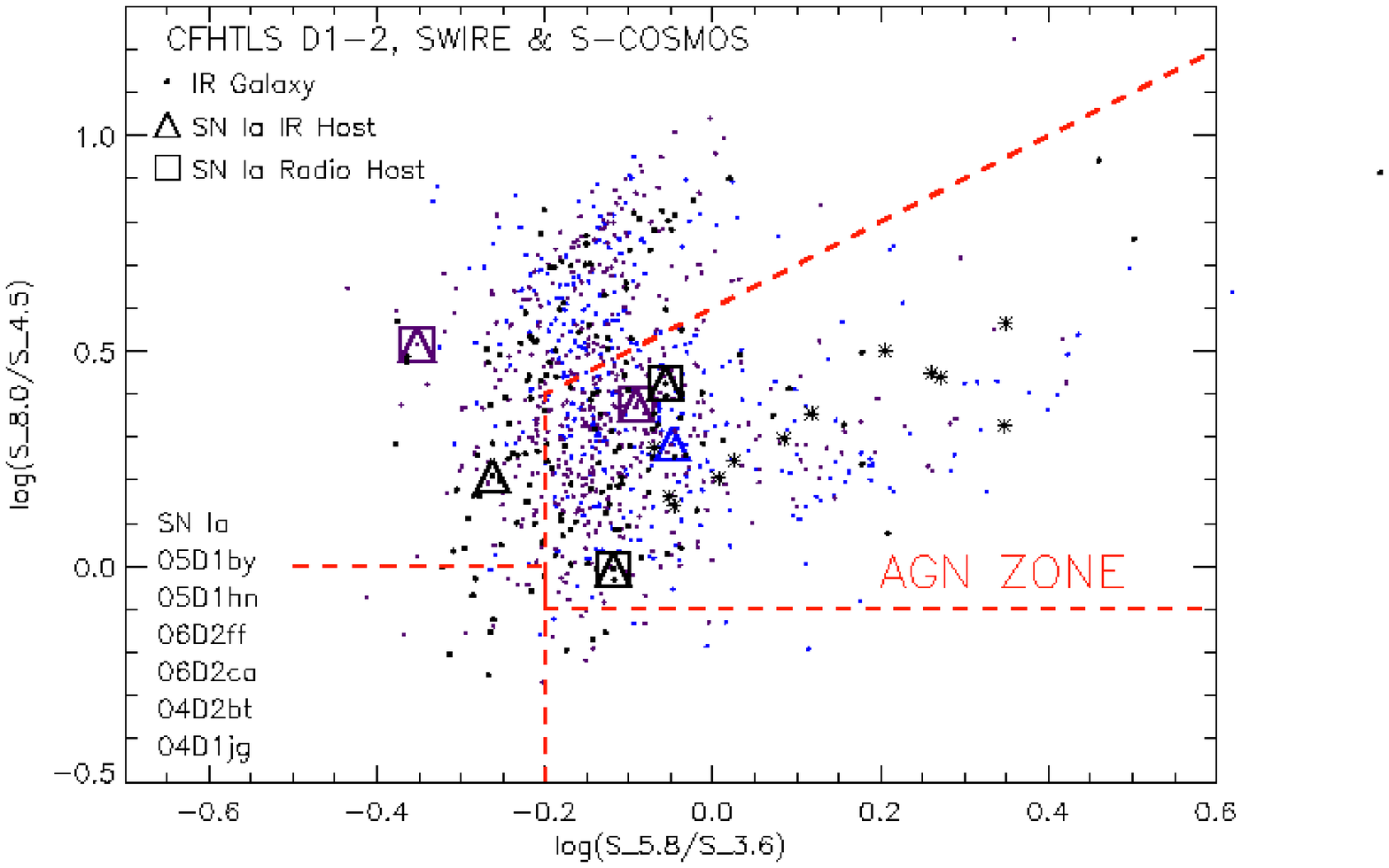}{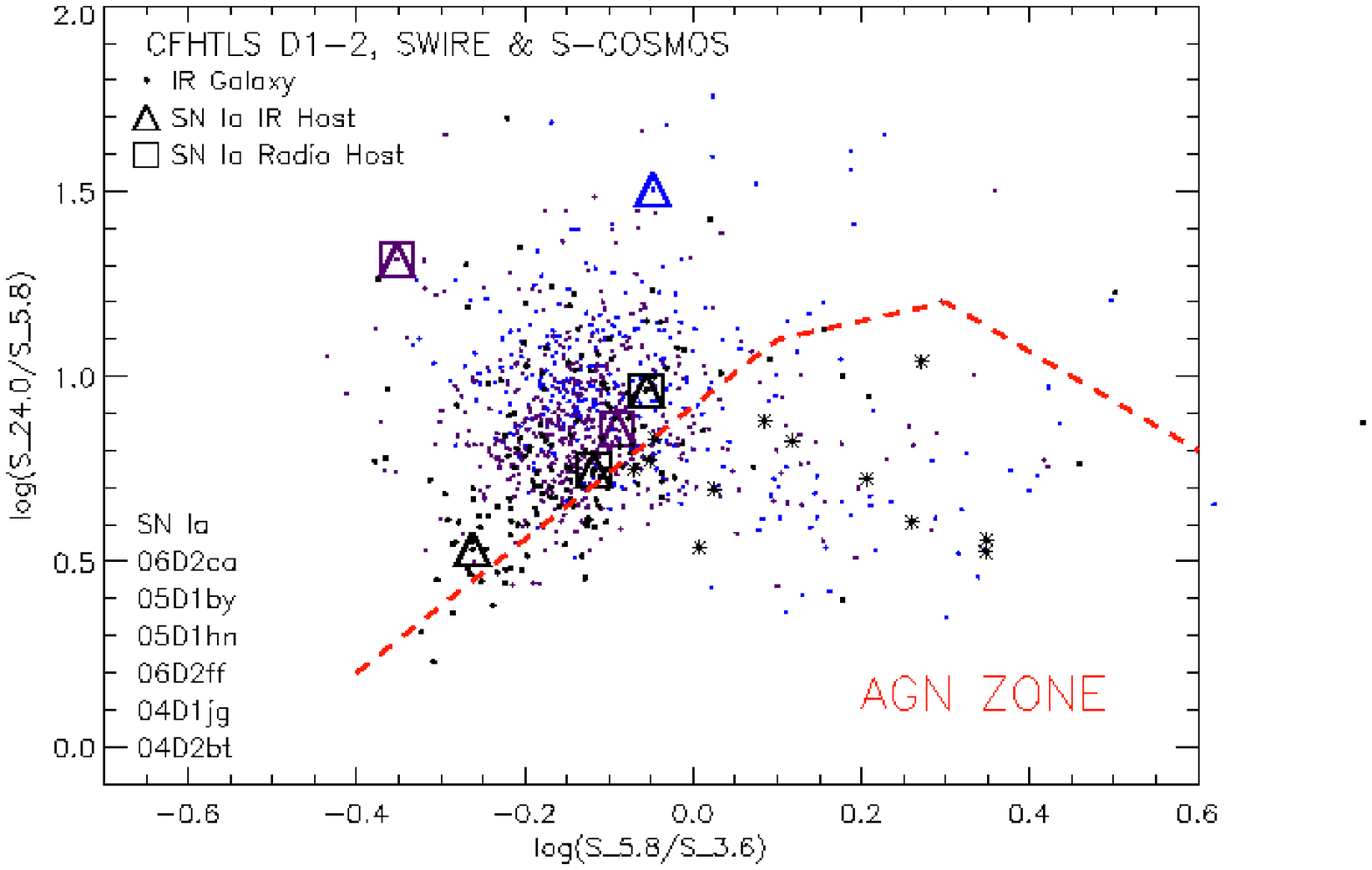}
      \caption{IR color-color diagrams for $\rm z\leq0.6$ galaxies with IRAC counterparts only (left)
        and IRAC+MIPS counterparts (right), for D1 and D2 combined. Axes are infrared colors, the
        logged ratios of fluxes in two bands (i.e. S\_3.6 is the flux at $\rm 3.6\ \mu m$).
        Filled circles are galaxies of SED type E/S0 (black), Sbc (purple), and later types (blue).
        Triangles are SN\,Ia host galaxies, with squares for radio hosts. Red dashed lines
        mark AGN boundaries as in Figure 10 of Sajina et al. (2005) and Figure 2 of Lacy et al. (2004).
        Asterisks mark E/S0 galaxies in {\it both} AGN zones. Lower left lists SN\,Ia
        in order of decreasing y-axis value. \label{FS2IRcc}}
    \end{center}
  \end{figure}

\begin{figure}
    \begin{center}
      \plottwo{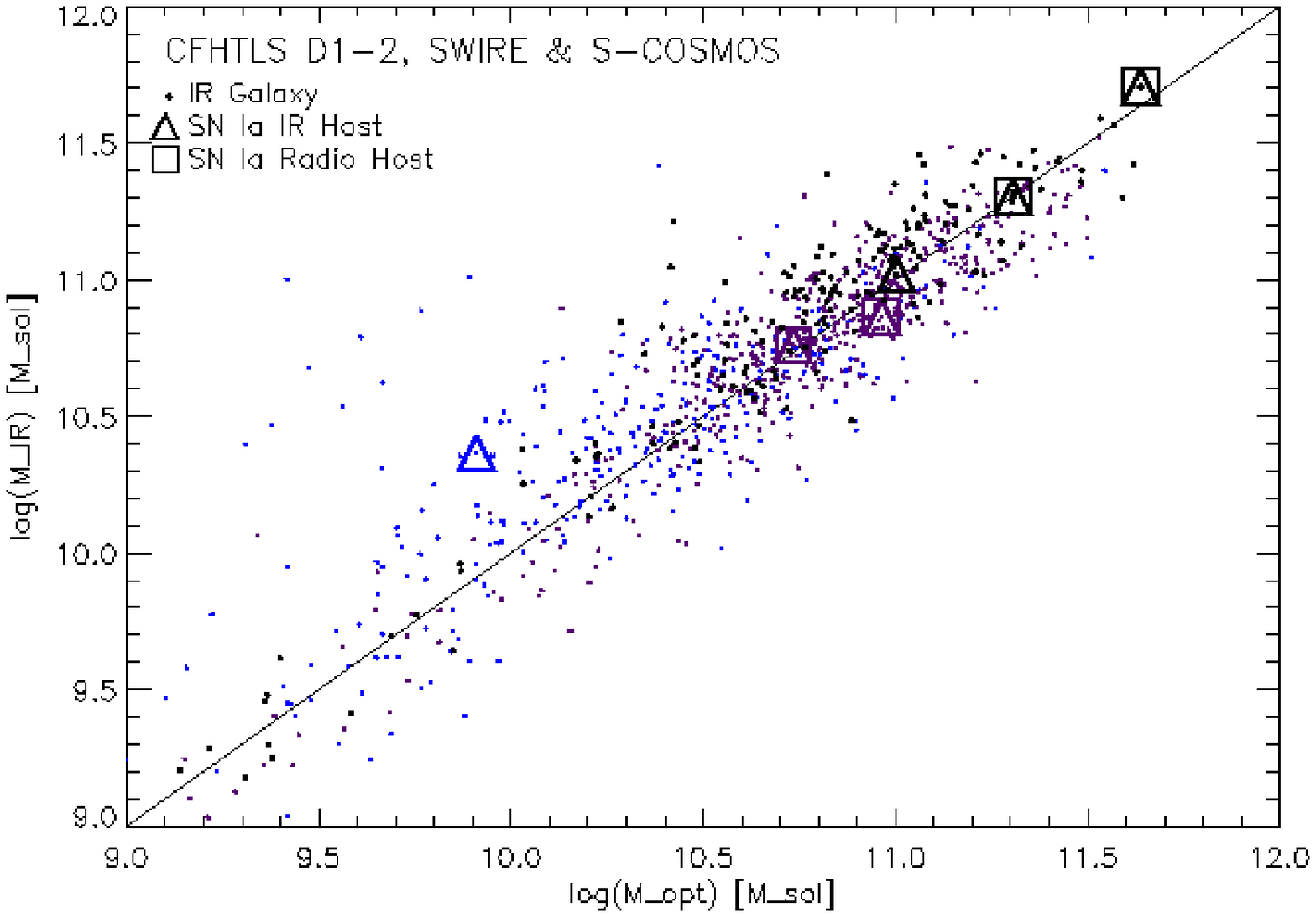}{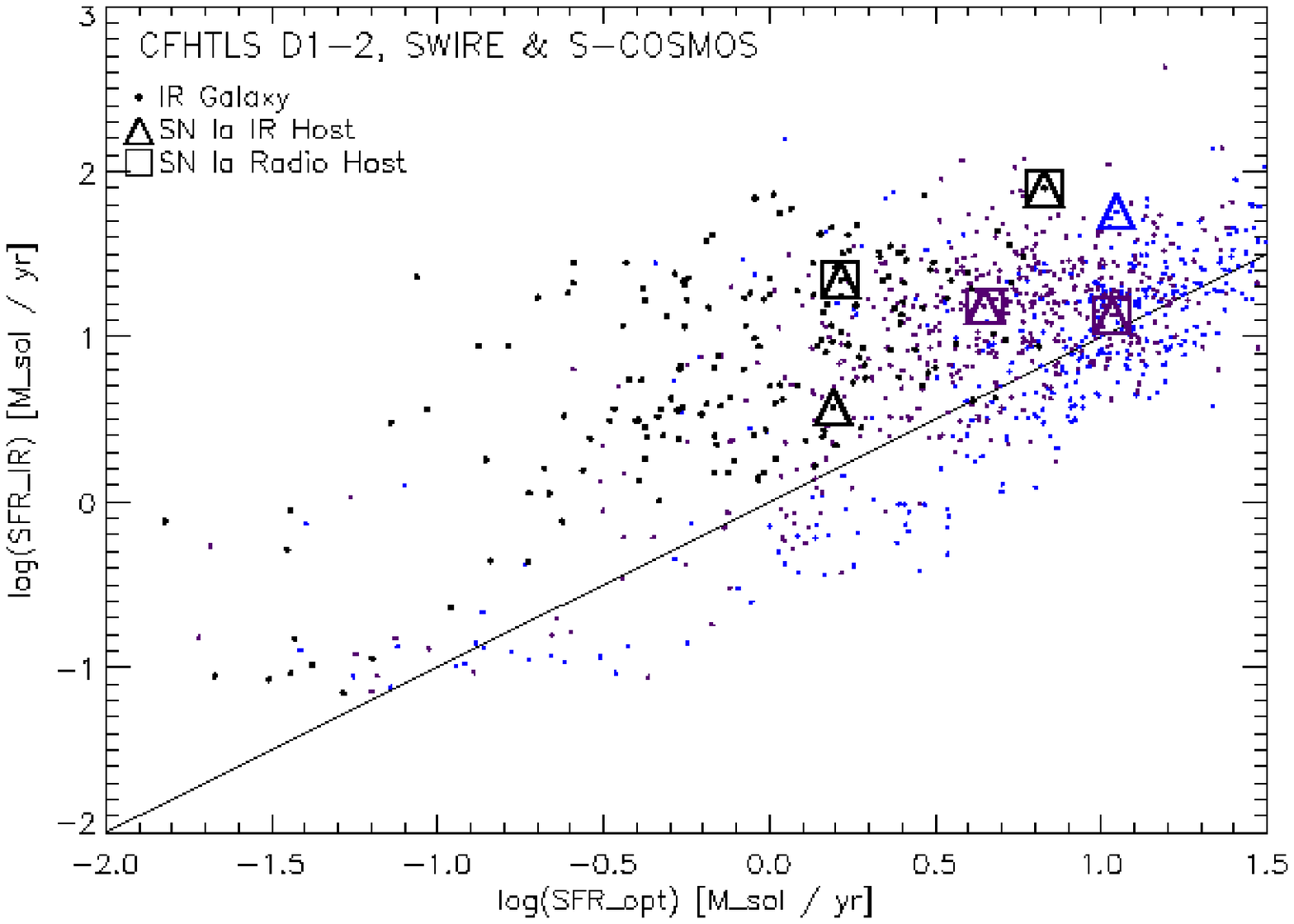}
      \caption{IR versus optical mass (left) and SFR (right) for $\rm z\leq0.6$ galaxies
        IRAC+MIPS counterparts, for D1 and D2 combined. Filled circles are galaxies
        of SED type E/S0 (black), Sbc (purple), and later types (blue).
        Triangles are SN\,Ia host galaxies, with squares for radio hosts. Solid
        lines of slope equal to one are there to guide the eye. \label{FS2OPTIRcomp}}
    \end{center}
  \end{figure}

  We convert $\rm S_{3.6\mu m}$ to K-band stellar mass via Balogh et al. (2007),
  and $\rm S_{24\mu m}$ to bolometric infrared luminosity $\rm L_{IR}$ and
  infrared star formation rate $\rm SFR_{IR}$ with publicly available templates and
  codes\footnote{http://david.elbaz3.free.fr/astro\_codes/chary\_elbaz.html} \citep{CE01}.
  Plots of $\rm L_{IR}$ versus redshift in Figure \ref{FS2IRlumVSz} show
  our sample of Luminous Infrared Galaxies (LIRG, $\rm L_{IR} > 10^{11} L_{\odot}$)
  is complete to $\rm z\sim0.5$.
  The conversion of $\rm L_{IR}$ to $\rm SFR_{IR}$ is appropriate for starburst-dominated infrared emission,
  but IR color-color diagrams in Figure \ref{FS2IRcc} show some IR sources may be AGN-dominated;
  uncertainties introduced by applying this conversion to potentially AGN-dominated sources are discussed in \S~\ref{sSABPir}.
  Figure \ref{FS2OPTIRcomp} shows the optical and infrared masses agree very well, but
  $\rm SFR_{opt}$ is 3--10 times lower than $\rm SFR_{IR}$. Having sampled the
  brightest infrared galaxies which are most likely to have obscured star formation, this factor is
  and should be higher than the usual $\rm SFR_{IR} \sim 2 \times SFR_{opt}$ \citep{Tak05}.


 \subsection{Properties of SNLS SN\,Ia Host Galaxies}
  \label{sS2hostradioIR}

\begin{figure}
    \begin{center}
      \includegraphics[width=4.0cm,height=4.0cm]{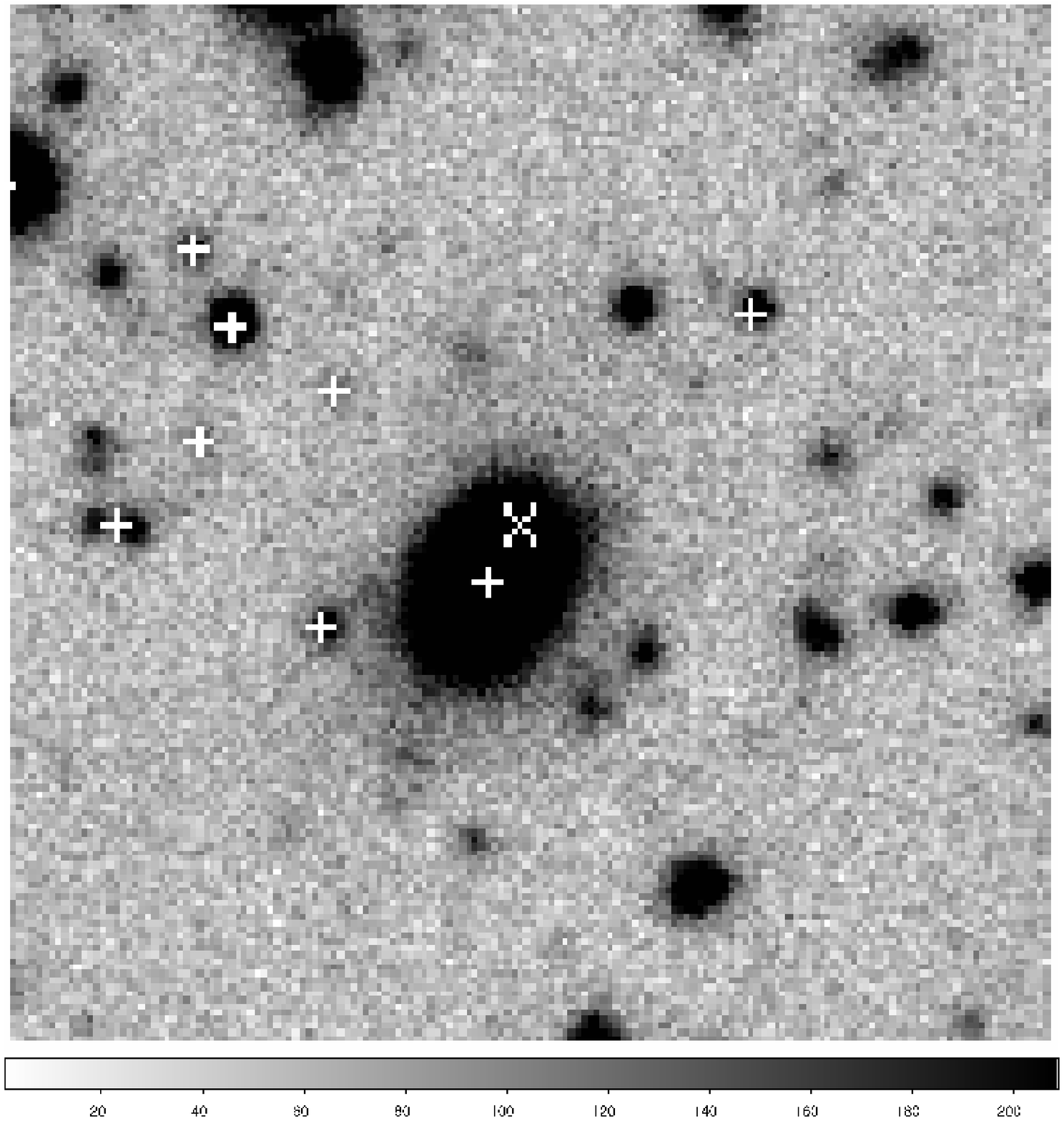}
      \includegraphics[width=4.0cm,height=4.0cm]{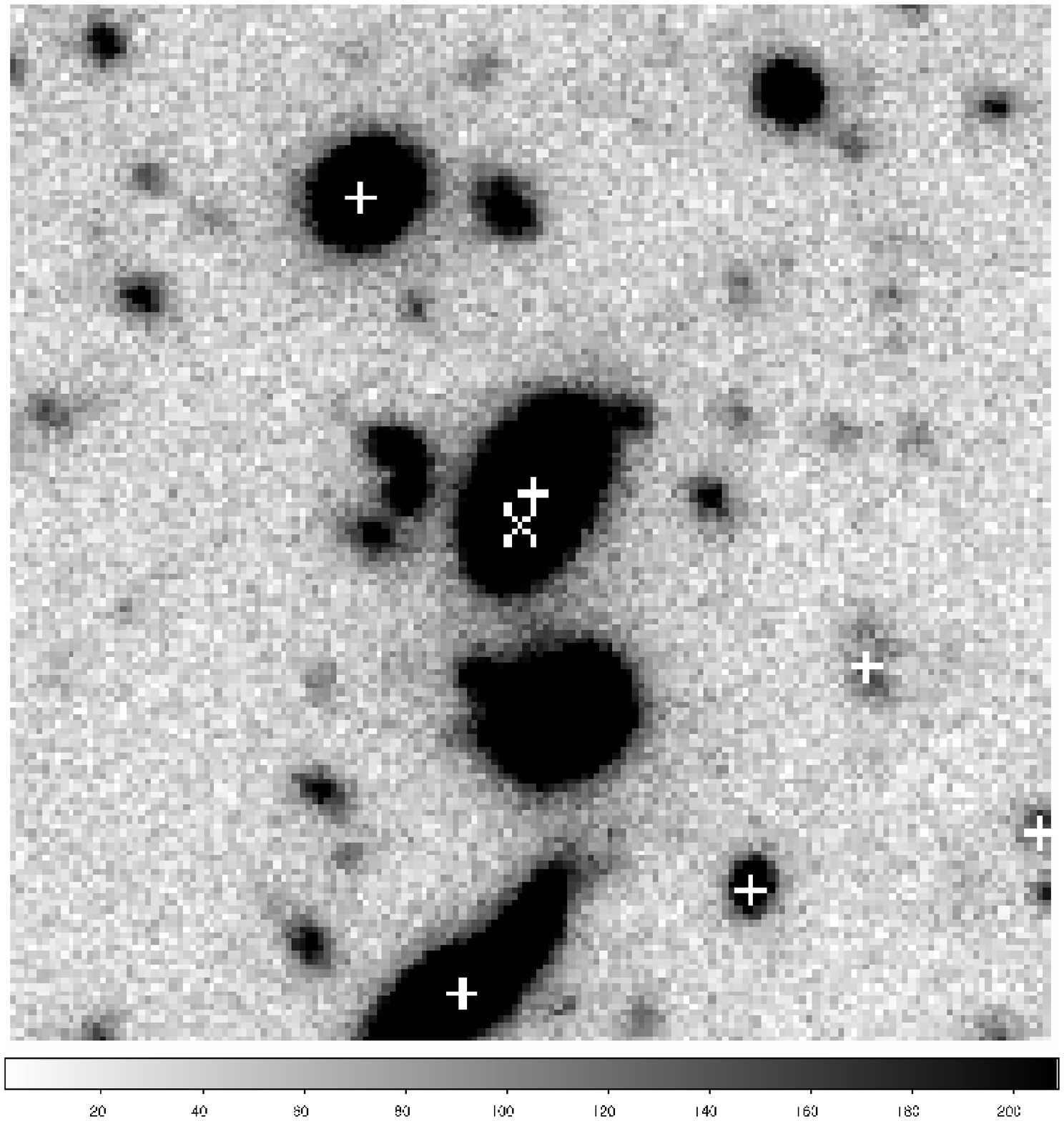}
      \includegraphics[width=4.0cm,height=4.0cm]{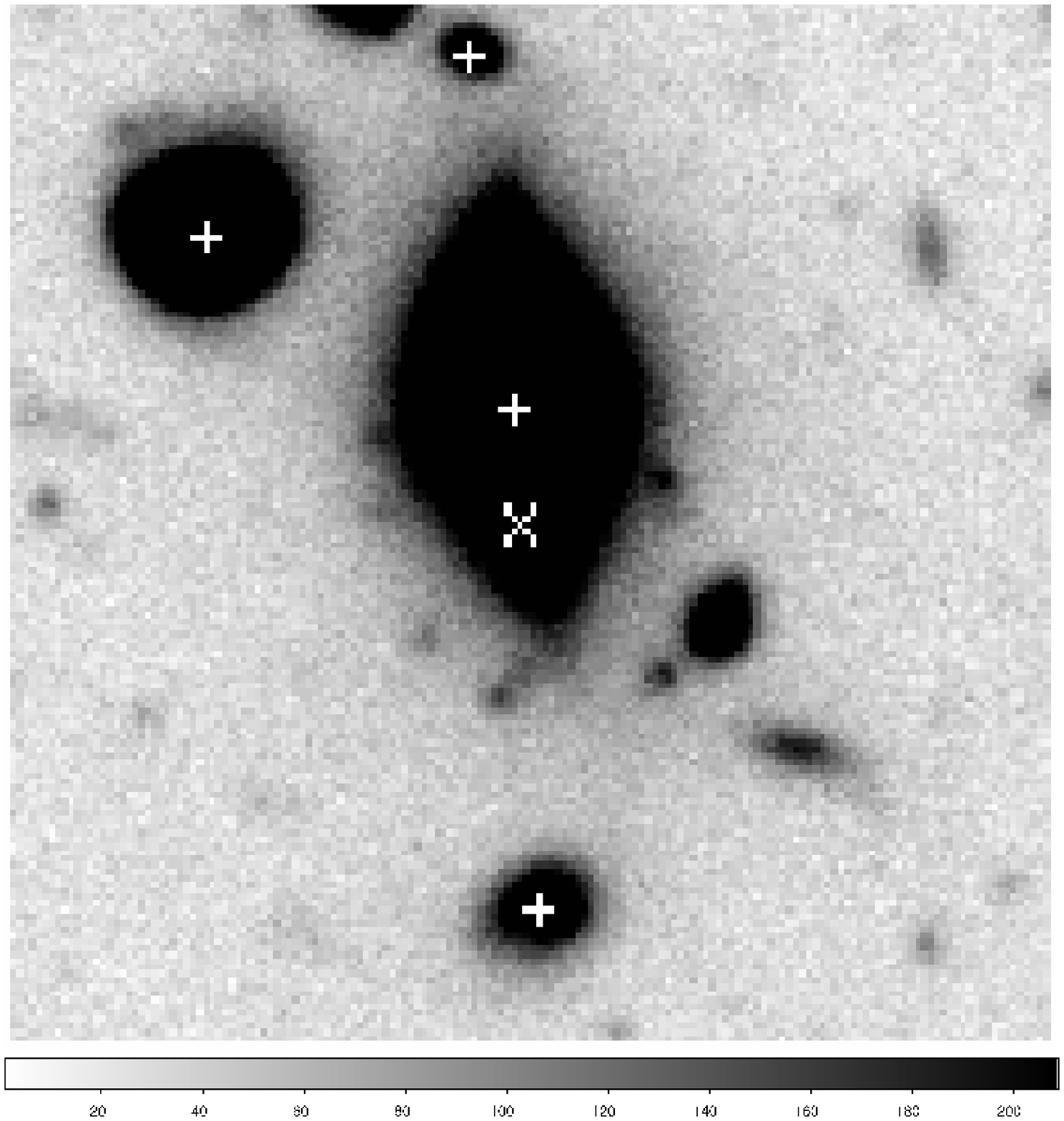}
      \includegraphics[width=4.0cm,height=4.0cm]{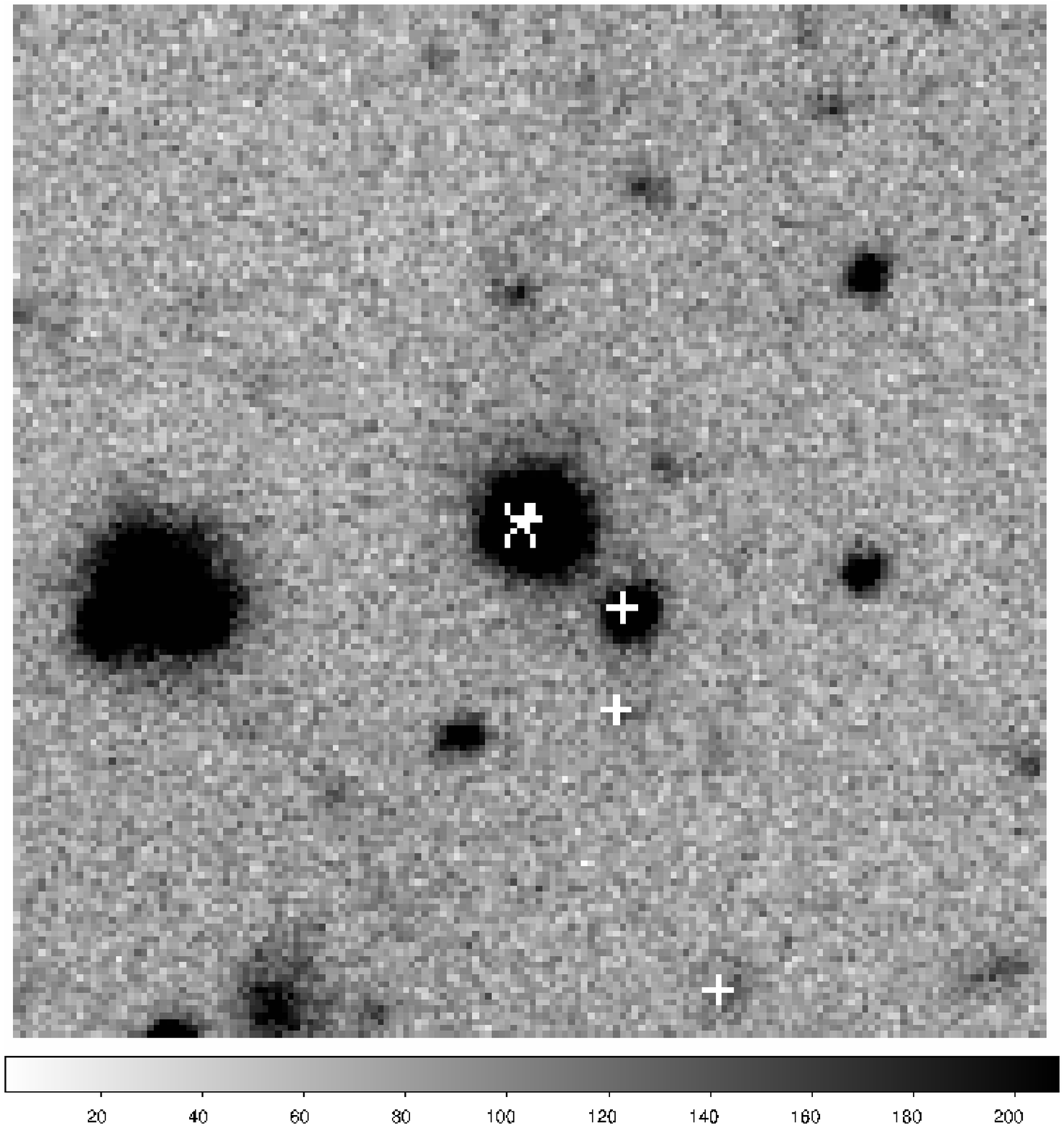}
      \includegraphics[width=4.0cm,height=4.0cm]{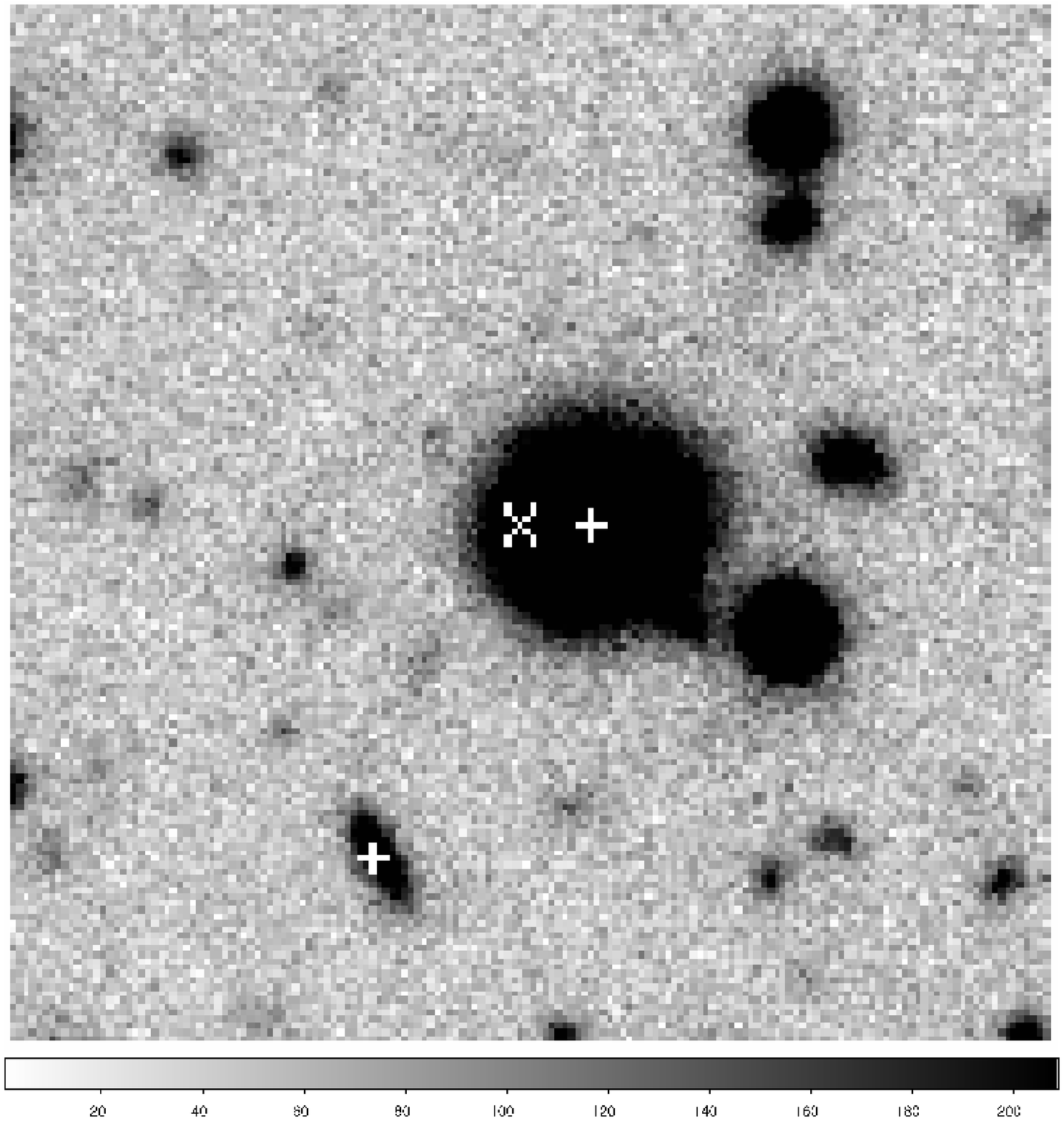}
      \includegraphics[width=4.0cm,height=4.0cm]{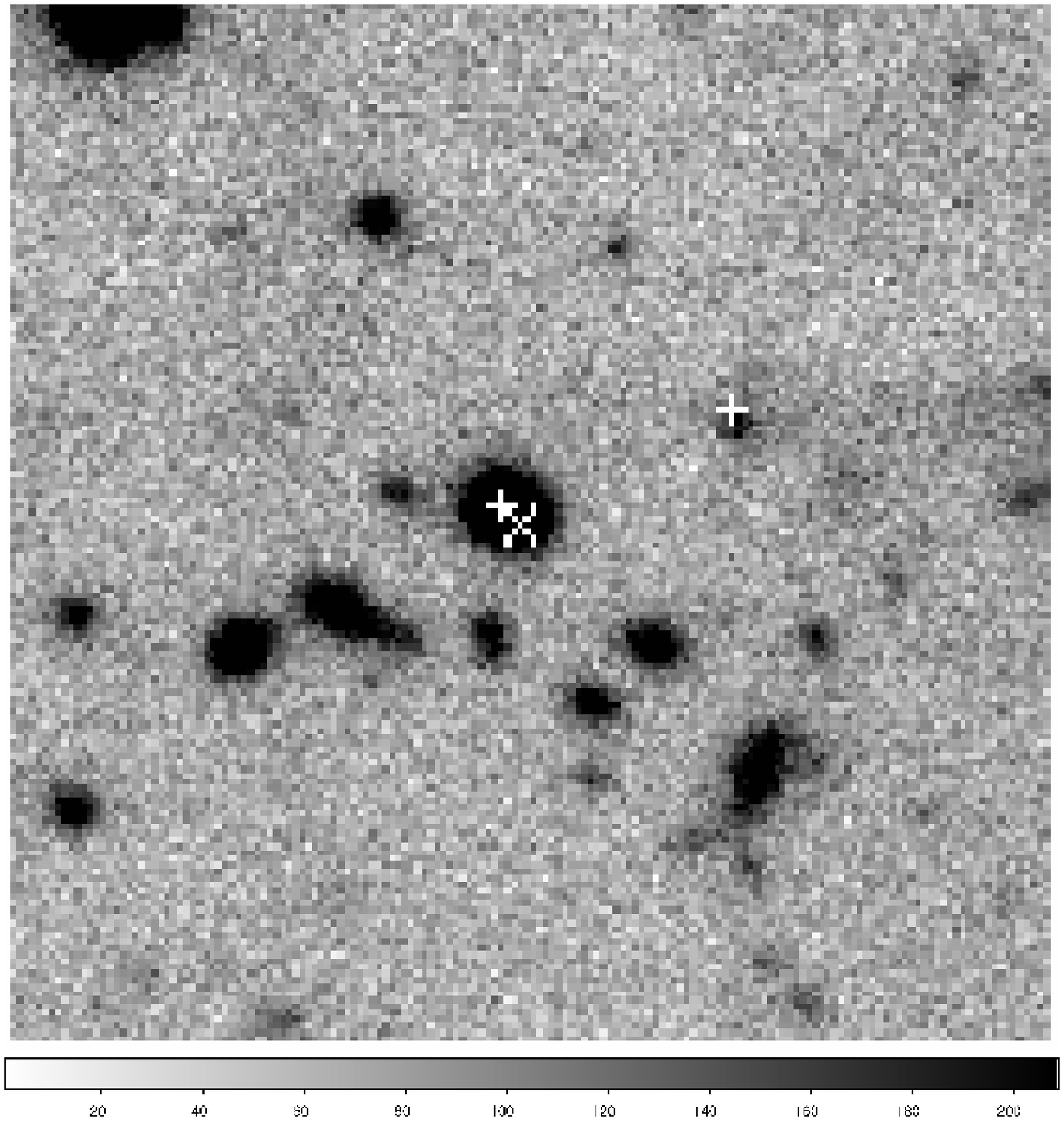}
      \includegraphics[width=4.0cm,height=4.0cm]{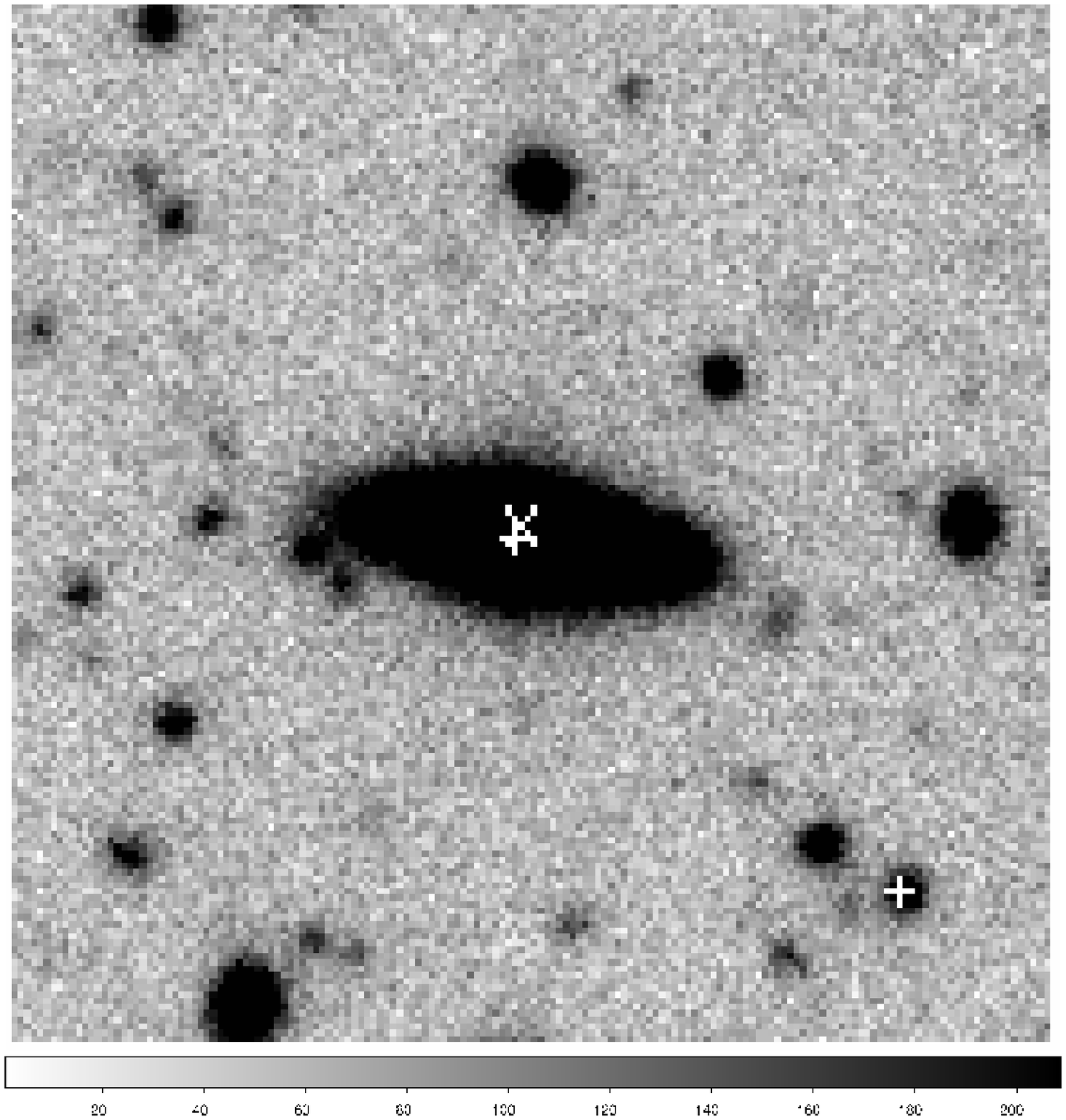}
      \includegraphics[width=4.0cm,height=4.0cm]{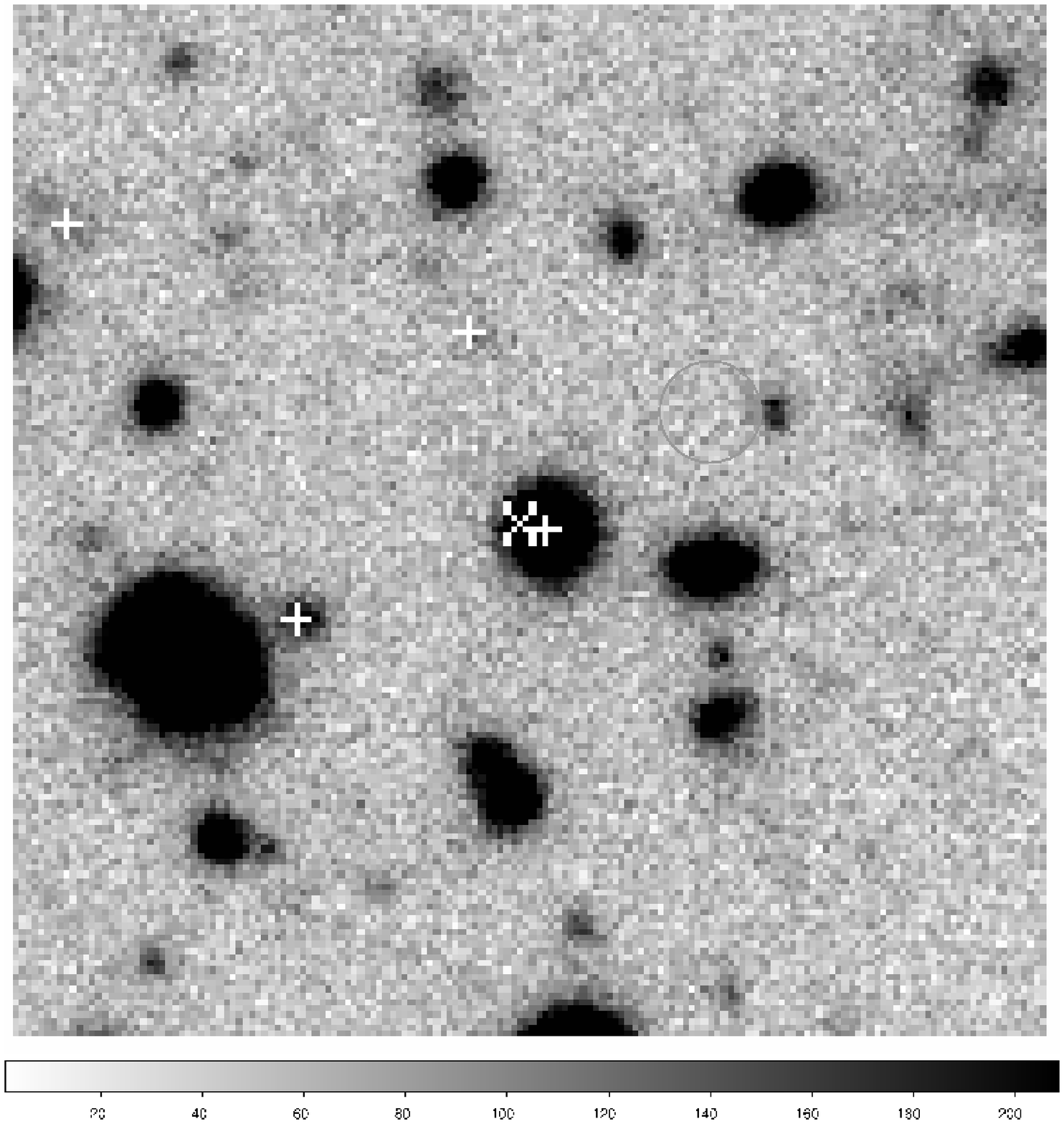}
      \caption{ Image stamps in $\rm r_{Mega}$, 30x30 arcseconds, centred on SNe\,Ia for host galaxies
        detected in radio and/or infrared source catalogs. Right to left, top to bottom: 04D1jg, 05D1by, 05D1hn,
        06D2ck, 06D2ff, 06D2je, 04D2bt, and 06D2ca.
        White crosses mark SNe\,Ia, and white plus signs mark nearby galaxies within $\rm \Delta z \leq 0.1$
        (i.e. $\rm \sim3\sigma_{\Delta z / (1+z)}$). \label{FS2images}}
    \end{center}
  \end{figure}

  We find eight SNLS SN\,Ia host galaxies with radio and/or
  infrared counterparts and $\rm z \lesssim 0.6$, the properties
  of which are given in Table \ref{TS2RLESSN} and plotted in
  Figures \ref{FS2RlumVSz} to \ref{FS2OPTIRcomp}. Image stamps
  centred on SN\,Ia coordinates are shown in Figure \ref{FS2images}.
  Beyond the redshift limit of $\rm z=0.6$, there are no radio SN\,Ia hosts,
  but one with an infrared counterpart: the Sbc type host of
  SN\,Ia 04D2ca at $\rm z=0.835$ is classified as a LIRG. In \S~\ref{Srelax}
  we discuss the effects of extending our redshift range to $\rm z\sim1.0$.
  Of these eight SN\,Ia hosts, five have SED type E/S0 and will be included
  in the analysis: four are radio-loud 
  ($\rm L_{1.4\,GHz} > 10^{29}\ ergs\ s^{-1}\ Hz^{-1}$), three are bright
  infrared galaxies (BIRG, $\rm L_{IR} > 10^{10} L_{\odot}$), and two are
  luminous infrared galaxies (LIRG, $\rm L_{IR} > 10^{11} L_{\odot}$). 

  Galaxies which are bright at mid-infrared wavelengths (MIR, for example $\rm 24 \mu m$) are 
  described as being starburst or AGN dominated, meaning they contain dust which
  absorbs high-energy photons from either young stars or AGN, and re-emits them in the MIR.
  It may seem the two radio-loud SN\,Ia host galaxies with SED type E/S0 (normally quiescent)
  and infrared counterparts must be AGN dominated, but there are several reasons why this
  is not the case. First, the radio-loud E/S0 hosts do not lie along the ``AGN plume''
  of the IRAC color-color plot (although 04D1jg and 05D1hn hosts are in its ``AGN zone''), and
  are outside the ``AGN zone'' of the IRAC+MIPS color-color plot as shown in Figure \ref{FS2IRcc}.
  Second, AGN-dominated infrared emission originates from a hot dusty torus
  tens of parsecs across \citep{Trist09} and would be unresolved, but the infrared
  sources associated with these two galaxies have are flagged as extended.
  Third, a visual inspection of radio and infrared SN\,Ia hosts in $u_M$ and GALEX (D2 only)
  images\footnote{http://www.cadc.hia.nrc.gc.ca/community/CFHTLS-SG/docs/cfhtls.html} show
  all but the late-type host of 06D2ff are quite faint to invisible in the ultraviolet,
  which can be indicative of dust obscured star formation.
  Fourth, not all starbursts appear morphologically irregular (e.g. 
  Figure 4.68 of Binney \& Merrifield (1998), in which a profoundly disturbed
  galaxy appears as an elliptical in a shallow exposure).
  We therefore accept the conversion of $\rm L_{IR}$ into
  $\rm SFR_{IR}$ for these radio-loud E/S0 SN\,Ia host galaxies, as discussed in \S~\ref{sS2IR}.  

  \subsection{Cappellaro et al. (1999) Catalog}
  \label{sS2C99}

  The Cappellaro et al. (1999) catalog is a combination of
  visual and photographic searches from Cappellaro et al. (1993) and
  Evans et al. (1989), provided to us by E. Cappellaro. SN data includes
  name, type (classified photometrically), and host name; galaxy catalog includes Hubble type,
  recession velocity, B-band luminosity and control time.
  For radio counterparts we use the same methods, matching criteria, flux limits,
  and radio source catalogs as DV05:
  the NRAO VLA Sky Survey \citep{NVSS}, the Parkes-MIT-NRAO survey \citep{PMN},
  and the GB6 survey \citep{GB6}. We convert radio flux to luminosity as in \S~\ref{sS2radio}.
  Table \ref{TDVE} shows we do not recover the same radio luminosities but
  do classify the same hosts as radio-loud, except for that of SN\,1983J
  which DV05 classify as ``borderline'' radio-loud and thus add 0.5 SN\,Ia
  to the number of radio-loud hosts. We find a lower radio power and would not do this.

  Galaxy classifications from the NASA Extragalactic Database\footnote{http://nedwww.ipac.caltech.edu} (NED)
  for the early-type SN\,Ia host sample of DV05 (their Table 2) are reproduced in Table \ref{TDVE},
  showing some are peculiar (possibly merging) and/or have emission lines.
  We derive mass and star formation rate from morphology and optical magnitude
  as in \S~\ref{sS2IlbCat}, and K-band stellar masses from the 2 Micron All-Sky Survey \citep{2MASS} as in \S~\ref{sS2IR}.
  We compile infrared data from two Infrared Science Archive \footnote{http://irsa.ipac.caltech.edu}
  catalogs, the IRAS Faint Source Catalog \citep{FSC} and the
  IRAS Cataloged Galaxies and Quasars \citep{galq}, and convert to infrared luminosity
  as described by Sanders \& Mirabel (1996).
  We also calculate the expected number of SN\,Ia for each galaxy from the two-component ``A+B''
  model as described in \ref{sS2add}, and use the given given galaxy control times
  (the time during which an exploding SN would be detectable)
  which also account for SNe detection biases \citep{C97}.
  Optical properties have an uncertainty of $\sim 40$--$60\%$ from conversions
  between galaxy type conventions, magnitude bands, and mass and SFR models.

  The properties of early-type C99 SN\,Ia hosts presented in Table \ref{TDVE}
  suggest they may not all be simply quiescent ellipticals, perhaps
  containing more SN\,Ia progenitors (of the `prompt' or `B' component variety)
  than previously expected. For example, NGC 1275 is a LIRG and likely contains dust-obscured
  star formation and a young stellar component despite being morphologically early-type.

  \section{SN\,Ia RATES IN RADIO-LOUD AND INFRARED-BRIGHT GALAXIES}
  \label{Srates}

  In this section we present SNLS and C99 SN\,Ia rates in a variety
  of early-type host galaxy subsets: radio-loud ($\rm L_{1.4\,GHz} > 10^{29}\ ergs\ s^{-1}\ Hz^{-1}$),
  radio-loudest ($\rm L_{1.4\,GHz} > 10^{30}\ ergs\ s^{-1}\ Hz^{-1}$),
  bright infrared galaxies or BIRG ($\rm L_{IR} > 10^{10} L_{\odot}$),
  and luminous infrared galaxies or LIRG ($\rm L_{IR} > 10^{11} L_{\odot}$).
  The number of SNLS SN\,Ia observed in D1--2 prior to 2006 Dec 21 for each subset,
  $\rm N_{obs}$, is shown in Table \ref{TNSN}. Also shown (in brackets) is $\rm N_{obs}/C$, where C is
  the effective control time from Equation \ref{EC} and accounts for SNLS detection efficiencies
  and observing seasons. Thus, $\rm N_{obs}/C$ is the number of SN\,Ia which explode
  per year of SNLS observations. Also quoted are Poisson uncertainties on $\rm N_{obs}/C$
  at the 0.84 ($1\sigma$) confidence level from Gehrels (1986).
  The amount of mass in galaxies in each subset is given in Table \ref{TMgal}.
  We calculate the specific SN\,Ia rate in each galaxy subset, $\rm sSNR_{Ia}$, via:
  
  \begin{equation}
    \label{ER}
    sSNR_{Ia} = \frac{N_{obs}/C}{\sum{[M/(1+z)]}}.
  \end{equation}
  
  \noindent
  The $\rm (1+z)^{-1}$ in the denominator is a time dilation factor for each galaxy, which converts
  the units for $\rm sSNR_{Ia}\ (SN\,y^{-1}\,M_{\odot}^{-1})$ from observed to rest-frame time.
  Specific SN\,Ia rates are given in Table \ref{TSNRates} in the
  commonly used ``supernovae unit'' SNuM, which is equal to one supernova per century per $\rm 10^{10}M_{\odot}$.
  Uncertainties are propagated from those in Table \ref{TNSN} but
  {\it do not include} the uncertainty in galaxy masses.
  Ratios of SN\,Ia rates in radio and IR subsets over all E/S0 galaxies are given in Table \ref{TSNRatios}.
  These tables also contain equivalent values for the C99 SN\,Ia and galaxy catalogs.
  We use radio luminosities derived in \S~\ref{sS2C99} to classify field galaxies,
  but initially use radio luminosities from DV05 to classify SN\,Ia hosts in order
  to make a direct comparison with their results. We revise our results with our own luminosities
  and discuss in \S~\ref{sSratesradio} and \ref{sSABPradio}.

  \subsection{Radio-Loud Galaxies}
  \label{sSratesradio}

  With SNLS data, we find the specific SN\,Ia rate in radio-loud early-type galaxies
  to be $0.102^{+0.103}_{-0.057}$ SNuM, within 1$\sigma$ of the C99 result in Table \ref{TSNRates}.
  Whereas DV05 quote a $\sim4$ times SN\,Ia rate enhancement in radio-loud over radio-quiet
  early-type galaxies, VLA radio catalog incompleteness means we can identify only
  half of the radio-loud population (\S~\ref{sS2radio}) and cannot isolate
  a radio-quiet set for the SNLS (see Figure \ref{FS2RlumVSz}). In lieu of comparing with a
  radio-quiet set we consider the ratio between radio-loud and {\it all} early-types as a
  lower limit, and find it consistent with no enhancement at the 1$\sigma$
  and 2$\sigma$ levels for the SNLS and C99 catalogs respectively, as shown in Table \ref{TSNRatios}.

  Can we correct our radio incompleteness, and does it affect our results?
  While we could use radio-loud galaxy mass functions to estimate the mass missing from
  our radio-loud sample, we could not know how many SN\,Ia this mass hosted,
  so cannot correct the SN\,Ia rates for radio incompleteness.
  Instead, we simply use the limited mass in radio-loud galaxies we {\it do} find,
  and the limited sample of SN\,Ia hosted by it, as a representative sample.
  Any SN\,Ia rate enhancement in radio-loud galaxies would also appear
  in this limited sample, assuming it is not proportional to $\rm L_{1.4\,GHz}$ (i.e. AGN power). 
  The SN\,Ia rate is not expected to be proportional to $\rm L_{1.4\,GHz}$ because
  AGN activity is not suspected of directly influencing the probability of an SN\,Ia explosion,
  as discussed in \S~\ref{S1Intro}; any evolution in the physical properties
  of AGN is not expected to influence our results either.
  Our radio incompleteness is thus unlikely to affect our results - they are
  a suitable comparison sample to C99.
  
  As an extra test, we present the specific SN\,Ia rates amongst the ``radio-loudest''
  ($\rm L_{1.4\,GHz} > 10^{30}\ ergs\ s^{-1}\ Hz^{-1}$) early-type galaxies
  in Table \ref{TSNRates}, and their ratio over all early-types in Table \ref{TSNRatios}.
  Unlike the radio-loud, the radio-loudest sample from SNLS catalogs is complete to $\rm z=0.6$.
  While the SNLS SN\,Ia rate in the radio-loudest early-type galaxies shows no increase over
  the rate in all early-types, the C99 catalog shows a $\sim4 \times$ enhancement which is
  twice that for the radio-loud sample. However, this is within 3$\sigma$ of a null result,
  and additional factors may artificially inflate this rate as discussed below.

  Although we do not recover precisely the same radio luminosities for C99 early-type
  hosts as DV05, we have initially used their classifications in order to directly
  compare with their results. As shown in Table \ref{TDVE}, we would place NGC 1316
  in the radio-loud subset but not the radio-loudest, and would not classify
  NGC 3106 as ``borderline'' radio-loud. Also, SN\,1968A in NGC 1275 is listed
  as type-I, not type-Ia. Although rare, core-collapse supernovae
  have been observed in early-type galaxies; this is often due to misclassification of 
  late-types as early-type, especially for galaxies with HI and radio emission \citep{Hako08,Bazin09}.
  NGC 1275 is a known star forming early-type galaxy, has the highest $\rm L_{IR}$ of
  all C99 early-type hosts (Table \ref{TDVE}), and is precisely the sort to
  potentially host a CC\,SN. Considering these factors, we have instead 
  5, 8, and 20 SN\,Ia hosts in the radio-loudest, radio-loud, and all early-type
  samples, lowering their rates to $0.123^{+0.083}_{-0.053}$, $0.085^{+0.042}_{-0.030}$,
  and $0.042^{+0.012}_{-0.009}$ SNuM respectively. Ratios for the radio-loudest
  and radio-loud categories drop to $2.9^{+3.3}_{-1.6}$ and $2.0^{+1.9}_{-1.0}$,
  bringing the null result within 2$\sigma$ and 1$\sigma$ respectively.

  \subsection{Bright and Luminous Infrared Galaxies}
  \label{sSratesir}

  With SNLS data, we find the specific SN\,Ia rate in early-type BIRG and LIRG 
  to be $0.168^{+0217}_{-0.107}$ and $0.249^{+0.469}_{-0.191}$ SNuM respectively.
  As presented in Table \ref{TSNRates} the C99 data results agree within 1$\sigma$.
  Again we consider the ratios between B/LIRG and {\it all} early-types as a lower limit
  on any possible enhancement, as given in Table \ref{TSNRatios}. It appears
  the SN\,Ia rate is a factor of 3--5 times higher in B/LIRG, but this
  is not more significant than 1$\sigma$ (2$\sigma$ for C99 BIRG).
  While we could have isolated all early-type non-LIRG for a ratio, the LIRG
  fraction is small enough that this would not be significantly different
  from the all early-types sample.
 
  As discussed in \S~\ref{sSratesradio}, SN\,1968A in early-type LIRG
  NGC 1275 may not have been a type Ia. Also, SN\,1983J and SN\,1991Q
  do not have a definitive classification. The scenario that these
  three are not type Ia drops the rate in C99 BIRG early-type
  galaxies to $0.041^{+0.095}_{-0.034}$ SNuM, consistent with the
  SN\,Ia rate in all early-type galaxies. We also note that although
  one SN\,Ia host is optically classified as starburst (see \S~\ref{sS2IlbCat}),
  it does not have an IR counterpart and cannot be included here.

  \section{COMPARISON TO THE TWO-COMPONENT ``A+B'' MODEL}
  \label{SABP}

  Specific SN\,Ia rates in subsets of radio and infrared early-type
  host galaxies are $\sim$1--5 times the rate in all early-type
  galaxies, and all enhancement have low significance (2$\sigma$ at most).
  In this section we test whether these rates are also consistent with the
  two-component ``A+B'' model, and whether the potential enhancements might
  simply be due to star formation providing additional SN\,Ia progenitors.
  To do this, we statistically compare the observed number of SN\,Ia in radio
  and infrared early-type galaxies to that predicted by the two-component
  ``A+B'' model. For the SNLS catalog, the total number of SN\,Ia predicted by ``A+B''
  over the survey's duration is $\rm N_{A+B} = C \times R_{Ia}$ for an
  individual galaxy, where $\rm R_{Ia}$ and $\rm C$ are explained in \S~\ref{sS2add}.
  For the C99 catalog, $\rm R_{Ia}$ is converted to $\rm N_{A+B}$ using given observational control times.
  Where $\rm N_{obs}$ is the number observed, the Poisson probability of observing 
  $\rm x=N_{obs}$ given an expected number $\rm \mu=N_{A+B}$ is expressed by \citep{Bev}:
  
  \begin{equation}
    \label{EPP}
    P_P(x;\mu) = \mu^{x} \frac{e^{-\mu}}{x!}.
  \end{equation}
  
  \noindent
  When $\rm N_{obs} > N_{A+B}$, the {\it summed} Poisson probability
  of observing $\rm x=N_{obs}$ {\it or more} is the integral of this
  from $\rm x=N_{obs}$ to $\rm x=\infty$; when $\rm N_{obs} < N_{A+B}$ the probability
  of observing $\rm x=N_{obs}$ {\it or less} is the integral from $\rm x=0$ to $\rm x=N_{obs}$.
  It assesses whether the observed number of SN\,Ia  is consistent
  with the ``A+B'' model in any given galaxy subset, as used for 
  early-type cluster galaxies in Graham et al. (2008).
  Summed probabilities $\rm \leq 0.05$ are considered statistically significant results.

  We calculate summed Poisson probabilities for the same SNLS and C99 radio and infrared galaxy
  subsets described in \S~\ref{Srates}.  Where {\it optical} masses and star formation rates
  are used in the ``A+B'' model, we refer to the total expected number of SN\,Ia as
  $\rm N_{A+B,opt}$ and the associated summed probability as $\rm P_{opt}$. We can also derive $\rm N_{A+B,IR}$ and $\rm P_{IR}$
  by using infrared SFR if a galaxy has a MIPS counterpart (using optical SFR if not),
  an account for contributions from dust-obscured star formation to the ``prompt'' component.
  When substituting $\rm SFR_{IR}$ into the ``A+B'' model in Equation \ref{EAB},
  we must alter the ``B'' value of Sullivan et al. (2006) because it was
  derived from {\it optical} star formation rates, and typically
  $\rm SFR_{IR} \approx 2\ SFR_{opt}$ \citep{Tak05}. Although Figure \ref{FS2OPTIRcomp}
  shows $\rm SFR_{IR}\sim 3$--$\rm 10 \rm \times SFR_{opt}$, this is for the small fraction of brightest
  infrared galaxies in which this factor is, in most cases, due to
  dust-obscured star formation. Had $\rm SFR_{IR}$ been used for all regular galaxies,
  the derived B value would be $\rm B_{IR}\sim B_{opt}/2$.

  \subsection{Radio-Loud Galaxies}
  \label{sSABPradio}

  Summed Poisson probabilities for SNLS radio-loud and radio-loudest early-type galaxy
  samples are $>0.05$ as in Table \ref{TPP}, indicating no significant deviation
  between observations and ``A+B'' model predictions. As discussed in \S~\ref{sSratesradio},
  it is unlikely the SNLS results are affected by our radio incompleteness.
  
  However, results for the C99 samples suggest significant to {\it very} significant
  deviations in the samples of radio-loud and radio-loudest early-type galaxies. These
  results are unchanged with the consideration of infrared SFR. Could this be due to a
  $\sim5\times$ variation between derived B values from different surveys \citep{Greggio08}?
  For the C99 radio-loud sample, using $5\times$B yields $\rm N_{A+B,opt}=5.9$ and $\rm P_{opt}=0.11$,
  but $20\times$B is necessary to raise $\rm N_{A+B,opt}\sim 4$ and $\rm P_{opt}\gtrsim 0.05$
  for the C99 radio-loudest early-type subset, so a change in B value cannot be the whole story.
  As discussed in \S~\ref{sSratesradio}, the actual number of SN\,Ia in
  the radio-loud and -loudest subsets of C99 may be 8 and 5 respectively. This would increase $\rm P_{opt}$
  and $\rm P_{IR}$ to $\gtrsim 0.1$ for both subsets, which we do not consider significant.

  \subsection{Bright and Luminous Infrared Galaxies}
  \label{sSABPir}

  Summed Poisson probabilities in Table \ref{TPP} for SNLS BIRG and LIRG
  early-type samples {\it hints} at a possible excess of SNe\,Ia over
  optical ``A+B'' model predictions, and the inclusion of $\rm SFR_{IR}$ does
  raise $\rm N_{A+B,IR}$ and $\rm P_{IR}$, but these results are not statistically
  significant. The same can be said for C99 BIRG, though as discussed in
  \S~\ref{sSratesradio} the number of C99 SN\,Ia in BIRGs may be just 1,
  in good agreement with $\rm N_{A+B,opt}$. In general we find the number of
  SN\,Ia in BIRG and LIRG is consistent with predictions of the ``A+B'' model.
  
  We note in \S~\ref{sS2IR} that infrared emission of potentially AGN-dominated
  galaxies has been converted to $\rm SFR_{IR}$ with a starburst-dominated template,
  thereby artificially inflating $\rm N_{exp,IR}$. Identifying galaxies which lay
  along the ``AGN plume'' (left plot, plume between $\rm -0.1 \lesssim log(S_{5.8}/S_{3.6}) \lesssim 0.5$)
  {\bf and} in the ``AGN Zone'' (right plot) of Figure \ref{FS2IRcc} as the most
  likely to be AGN, we find this population contributes 20--40\% {\it of the additional}
  SN\,Ia predicted when $\rm SFR_{IR}$ is included, as in 20--40\% of ($\rm N_{A+B,IR}$)-($\rm N_{A+B,opt}$).

  \section{ALTERING THE DATA CONSTRAINTS}
  \label{Srelax}

  To consider whether altering the constraints on SNLS
  data effects our results we try, in turn: extending our redshift range
  to $\rm z=1.0$, including galaxies with SED type Sbc, and
  different combination methods for our radio catalogs. 

  Sixteen SNLS SNe\,Ia early-type host galaxies have $\rm 0.6<z<1.0$, but 
  due to the flux limits of our radio catalogs, none
  are associated with a radio or infrared counterpart (only the
  Sbc type host of SNLS SNe\,Ia 04D2ca at $\rm z=0.835$ is detected as a LIRG).
  We determine the effective control time $\rm C$ for D1 and D2 in
  redshift ranges 0.6--0.8 ($\rm C_1=0.897$ and $\rm C_2=0.298$) and 0.8--1.0 ($\rm C_1=0.369$ and $\rm C_2=0.134$)
  to account for varying SNLS detection efficiencies. The specific
  SN\,Ia rate in all early-type galaxies to $\rm z=1.0$ is $\sim0.054$ SNuM, not
  significantly different from that for $\rm z<0.6$. Extending
  to $\rm z=1.0$ adds mass - but not SN\,Ia - to the radio and infrared early-type
  subsets, reducing rates and ratios to insignificant levels
  and increasing summed Poisson probabilities to $\rm P>0.10$.
  No new information is gained from extending to $\rm z=1.0$ as radio
  and IR catalogs are not deep enough, and doing so only dilutes our sample.

  There are 34 galaxies with SED type E/S0 to Sbc and $\rm z<0.6$,
  6 radio-loud and 5 BIRG as in Table \ref{TS2RLESSN}.
  The specific SN\,Ia rate amongst all E/S0--Sbc galaxies is $\rm 0.061^{+0.015}_{-0.012}$ SNuM,
  (higher than for all E/S0); in the radio-loud set it is $\rm 0.101^{+0.077}_{-0.047}$ SNum (similar)
  and in the BIRG set is $\rm 0.081^{+0.071}_{-0.041}$ SNuM (lower). 
  No statistically significant SN\,Ia rate enhancement in radio or infrared over
  all E/S0--Sbc galaxies is observed, and in these sets the summed Poisson
  probability is $\rm P\gg0.05$. Including Sbc type galaxies adds to
  our total SN\,Ia in radio and infrared hosts, but their naturally
  larger B component makes it harder to tell whether any contribution to the B component
  is being made from star formation associated with radio and infrared emission. 
  
  The deeper flux limit of the VLA-COSMOS in D2 may result in slightly
  different radio source populations between D1 and D2. To investigate
  whether this effects our results, we first restrict VLA-COSMOS to
  the same flux limit as VLA-VIRMOS, $\rm S_{1.4GHz} > 0.08\ mJy$.
  Only 2 SN\,Ia in radio-loud early-type galaxies are observed, which
  decreases the specific SN\,Ia rate to $\sim 0.055$ SNuM, increases
  its ratio over all early-types to $\sim 1$, and raises the summed
  Poisson probability to $\rm P\sim0.5$. Secondly, we limit our analysis
  to the deeper radio catalog in D2 and are left with one radio-loud
  E/S0 SN\,Ia host, which yields a rate of $\sim 0.085$ SNuM, a ratio
  over all E/S0 $\sim 1.7$, and $\rm P=0.38$.
  In general we conclude that any differences in the radio source populations are not major
  contributors to the results of this experiment.

  \section{SN\,Ia PROPERTIES}
  \label{SSNProp}

  It is well established that SN\,Ia light curve shape and peak luminosity 
  are correlated: more slowly-declining light curves reach a brighter peak luminosity.
  Light curves are parametrized by $\rm \Delta m_{15}$, the decrease in magnitude over the
  first 15 days after maximum light \citep{Phil93}, or by stretch, $s$, the amount
  that a template (average) light curve must be ``stretched'' to fit the observations
  where $s>1$ for brighter, slower declining SNe\,Ia \citep{Perl97}. 
  SN\,Ia peak luminosity depends on the mass of $^{56}$Ni synthesized during the
  explosion \citep{Arnett82}; whether this mass is influenced more by the metallicity
  or age of the white dwarf progenitor star is under much debate \citep{TBT03,RH04,Gall05,Gall08,H09}.

  SN\,Ia properties are also correlated with qualities of the host galaxies,
  where late-type star-forming galaxies host mostly bright, slowly declining SNe\,Ia,
  and early-type elliptical galaxies with little to no star-formation host faint, rapidly declining
  SNe\,Ia \citep{Ham96,H01}. Gallagher et al. (2005) compiled a database of 57 local SNe\,Ia
  and found the mean $\rm \Delta m_{15}$ in late-type (Sa to Peculiar) host galaxies
  was $\rm \overline{\Delta m_{15}} \sim 1.1$, but in early-type (E/S0) hosts
  $\rm \overline{\Delta m_{15}} \sim 1.45$. Sullivan et al. (2006) show the median
  stretch of SNLS SNe\,Ia in late-type star-forming galaxies is greater than in early-types.
  This is supported by Howell et al. (2007) who use light curve fitting routine SiFTO \citep{Con08}
  to constrain the stretch distribution of ``B'' (prompt) component SNe\,Ia 
  (associated with young stellar populations) to be centred at $s=1.071$ with $\sigma=0.063$,
  and of ``A'' (delayed) component SNe\,Ia (associated with old stellar populations)
  to be centred at $s=0.945$ with $\sigma=0.077$.

  If SNe\,Ia in radio-loud early-type hosts are associated with young stars,
  their stretch values should be consistent with the ``B'' component stretch distribution.
  SiFTO stretch values for SNLS SNe\,Ia in radio galaxies are given in Table \ref{TS2RLESSN}.
  (SN\,Ia 06D2je was not included in the SNLS third-year results, but preliminary fits show its
  stretch is similar to 05D1hn). Stretches for SN\,Ia 04D1jg, 05D1hn, and 06D2ck
  are consistent with the ``B'' component distribution of Howell et al. (2007).
  Individually, 05D1hn and 06D2ck are more consistent with originating from the ``A'' component
  at the 1$\sigma$ level, but collectively these three have a mean stretch of $s=1.052$
  with $\sigma=$0.057, which overlaps the 1$\sigma$ uncertainty of the ``A'' component stretch distribution.

  Amongst C99 SNe\,Ia in radio-loud early-type hosts, five of eight have a $\rm \Delta m_{15}$
  given in Table 2 of DV05, $\rm \Delta m_{15}=\{1.28,1.33,1.73,1.88,1.30\}$. Their mean is
  $\rm \overline{\Delta m_{15}}=1.5$ with $\rm \sigma_{\Delta m_{15}}=0.25$ which is
  consistent with $\rm \overline{\Delta m_{15}} \sim 1.45$ for SNe\,Ia in early-type
  hosts \citep{Gall05}. Instead of a numerical comparison,
  DV05 find that the distribution of decline rates for SNe\,Ia in radio-loud early-type  
  hosts is intermediate between that found for late-type and early-type hosts.
  They suggest this supports a {\it continuum} of SN\,Ia and host galaxies properties,
  from young progenitors in very active galaxies, intermediate age progenitors in active
  galaxies, and old progenitors in passive galaxies. Stretch values for the three SNLS SNe\,Ia
  in radio-loud early-type galaxies also support this.

  Two SNLS SNe\,Ia found in radio-loud {\it late}-type hosts, 05D1by and 06D2ff,
  have stretch values consistent with the ``B'' and ``A'' component distributions
  respectively which is normal since late-types have an underlying old stellar population.
  Two SNLS SNe\,Ia in non-radio BIRG hosts, 04D2bt (early-type) and 06D2ca (late-type),
  have stretch values consistent with the ``A'' and ``B'' component distributions which
  also agree with their host type.

  In general we find the properties of radio and infrared SN\,Ia hosts, and their
  environments, to resemble dusty starburst galaxies (\S~\ref{sS2hostradioIR}),
  and expect their SN\,Ia light curves to be reddened and extinguished.
  SNLS analysis finds 05D1by, 05D1hn, 06D2ff, and 04D2bt are particularly red,
  and the first three have large Hubble residuals, indicating this is an unusually
  red, faint sample of SNe\,Ia.
  It is interesting that 04D1jg and 06D2ca are not particularly red or faint,
  because these two SN\,Ia have the {\it brightest} infrared hosts. Perhaps they
  originated on the 'near' side of the galaxy and escaped dust extinction
  (neither have large host offsets, 2 and 0.5 arcseconds respectively).
  Also, the host of 06D2ca is SED type Scd and likely has unobscured star forming
  regions. 
  
  As a final note, SN\,Ia 05D1by is also spectroscopically peculiar in that the
  $\rm 6150\AA$ Silicon line is very broad, indicating fast ejecta velocities.
  It is similar to SN 2001ay \citep{ayNug}, which had an atypically slow
  light curve decline given its unremarkable peak magnitude, a very luminous near-IR
  magnitude at $\sim10$ days post maximum, and an early-type host\citep{ayPhil}.
  Such peculiar spectra comprise $\sim1\%$ of all SNe\,Ia, but in this small sample
  it is impossible to comment on whether they are more or less common in active galaxies.

  \section{DISCUSSION}
  \label{SDisc}

  With the C99 and SNLS catalogs we recover the SN\,Ia rate in radio-loud
  early-type galaxies of DV05, and for the first time we calculate SN\,Ia
  rates in the radio-loudest, BIRG, and LIRG subsets of early-type galaxies.
  We find these rates are $\sim$1--5 times the rate in all early-type galaxies,
  at that any potential enhancement is $\lesssim 2\sigma$. For the first time,
  we have incorporated infrared star formation rates into the two-component
  ``A+B'' model. While it does slightly increase the number of SN\,Ia predicted,
  more so in B/LIRG than in radio-loud(est) samples, ultimately {\it all} SNLS
  results are consistent with the two-component ``A+B'' model. The same cannot
  initially be said for all C99 results, but the observed number of C99 SNe\,Ia
  may be an upper limit due to SN\,Ia and/or host misclassifications.

\begin{figure}
    \begin{center}
      \plottwo{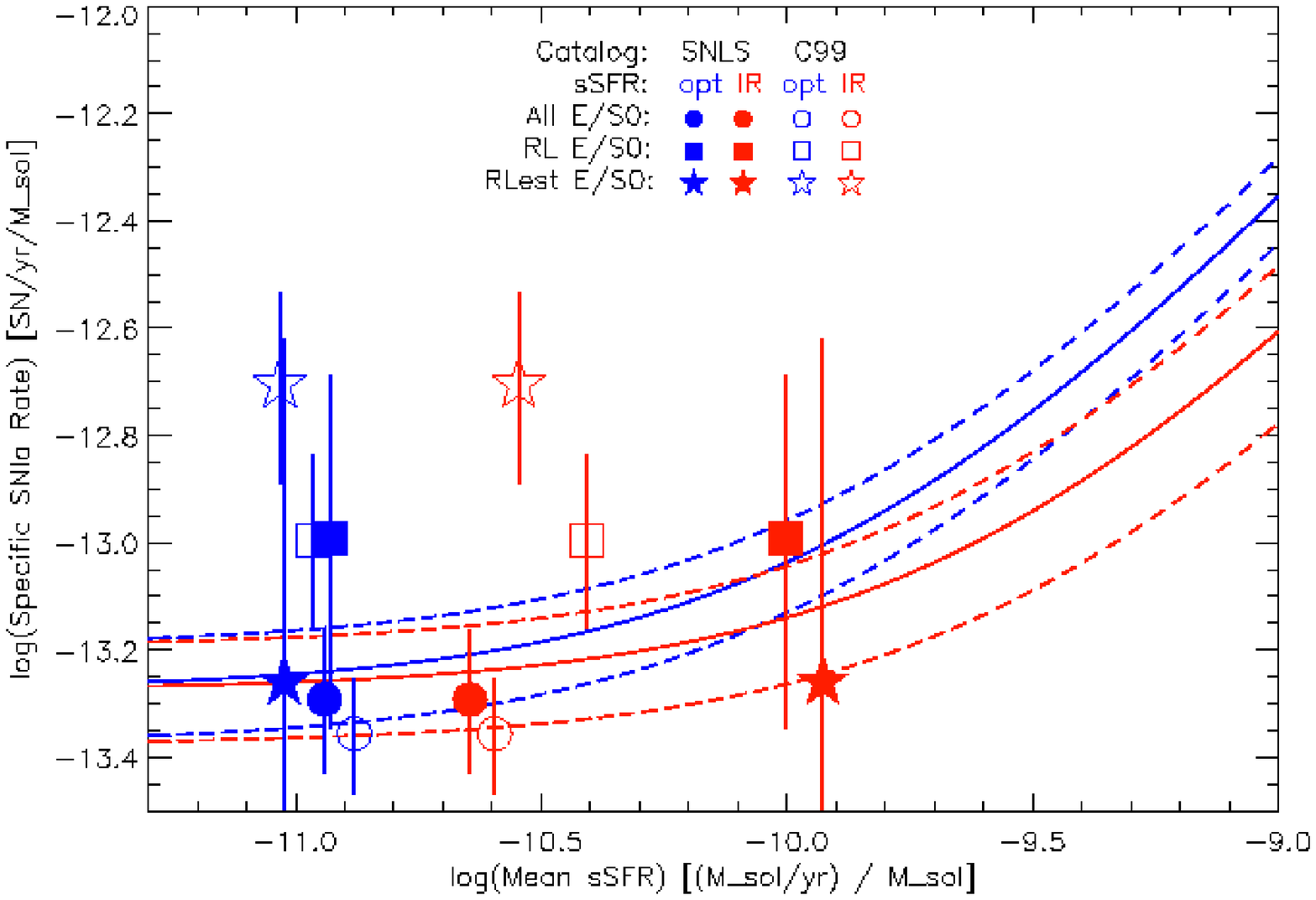}{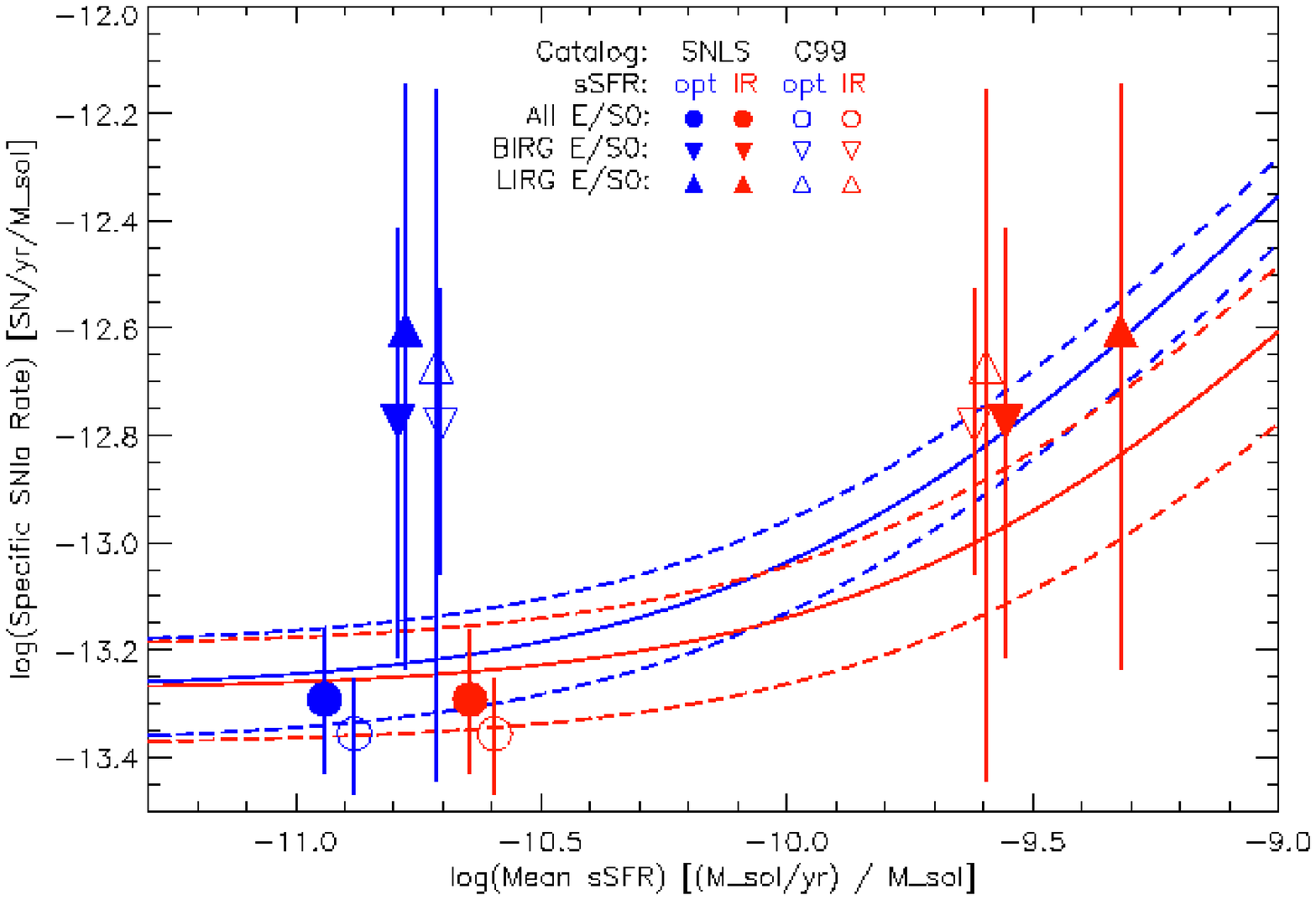}
      \caption{Specific SN\,Ia rates in SNLS (filled) and C99 (open) early-type galaxies (circles) from
        Table \ref{TSNRates}, versus mean sSFR of each subset. Left plot shows radio-loud (RL, square)
        and radio-loudest (RLest, star) subsets; right plot shows BIRG (inverted triangle) and
        LIRG (triangle) subsets. Symbol color indicates sSFR derived from optical data only (blue),
        or by incorporating infrared data when available (red). Also plotted is the two-component
        ``A+B'' model (solid) with uncertainties (dashed) from Sullivan et al. (2006) (blue lines),
        and the theoretical infrared ``A+B'' model (with $\rm B_{IR}\sim B_{opt}/2$) (red lines)
        as discussed in \S~\ref{SABP}.       \label{Frates}}
    \end{center}
\end{figure}

  The benefit of considering infrared host galaxy properties to account for possible dust obscured
  star formation is best shown in Figure \ref{Frates}, which plots the specific SN\,Ia rate as a
  function of mean specific star formation rate (sSFR) for each early-type galaxy subset.
  Blue symbols indicate where sSFR was derived from optical data only, and red symbols where
  $\rm SFR_{IR}$ was substituted for all galaxies with counterparts in the Spitzer IR catalogs.
  Curved lines show the two-component ``A+B'' model from Sullivan et al. (2006).
  Although error bars are large and all rates are within $2\sigma$ of the optical ``A+B'' model,
  when plotted with $\rm sSFR_{IR}$ the SNLS SN\,Ia rates in B/LIRG and radio-loud(est) subsets
  do align more closely with ``A+B''. The same can be said for C99 B/LIRG, and the apparent
  discrepancy for the C99 radio-loud(est) subset may be due to misclassifications discussed
  in \S~\ref{sSratesradio}. These results are not significant, but are included to demonstrate
  the utility of such an analysis for future SN\,Ia rates.

  DV05 suggested the SN\,Ia rate enhancement in radio-loud early-type galaxies
  may be due to galaxy mergers triggering radio activity and providing extra
  SN\,Ia progenitors by induced star formation. Some aspects of the SNLS data
  are consistent with this: the evidence for dust-obscured star formation in
  SN\,Ia radio hosts, the agreement between observations and ``A+B'' model
  predictions in SN\,Ia radio hosts, and the properties of SN\,Ia in radio
  hosts being similar to those associated with intermediate age stellar populations.
  However, the discrepancy between observations and ``A+B'' model predictions for C99 radio
  galaxies remains. In \S~\ref{sSratesradio} and \S~\ref{sSABPradio} we discuss
  possible SN\,Ia and/or host misclassifications, but might it be the two-component model
  is simply inappropriate for radio galaxies? It is a linear
  approximation to a relation that is not necessarily linear, and uses current SFR
  as a measure of the SFR when the white dwarf progenitor star was born - the parameter
  actually related to the number of SN\,Ia progenitors. In regular spiral galaxies where
  the SFR remains constant for $10^{9}$ to $10^{10}$ Gyr this is essentially true,
  but not necessarily for galaxies experiencing interactions and/or mergers, and perhaps
  episodic star formation (discussed below), on timescales of $\sim10^8$ years (DV05).
  This is not a failure of the ``A+B'' model, merely a limitation of its application to
  individual galaxies. In large samples this effect should approximately ``average out''.

  With such a hypothetical episodic SFR, Mannucci et al. (2006) show the radio-loud
  SN\,Ia rate enhancement of DV05 is best fit with a {\it bimodal} delay-time
  distribution (DTD) for SN\,Ia in which $\sim50\%$ of the prompt component explodes
  in $\lesssim10^8$ years, and suggest two physically distinct populations of SN\,Ia
  progenitors. However, we find the SNLS SN\,Ia rate in radio galaxies agrees with
  the ``A+B'' model, which is consistent with {\it continuous} DTD's \citep{Pr08},
  and that the properties of SNe\,Ia in radio hosts do not constrain them to very
  young stellar populations (as does DV05).
  
  As a final note, the suggestion by DV05 that galaxy interactions and/or mergers cause both
  the radio activity and enhanced SN\,Ia rate is consistent with the well documented
  correlation between clustered environments and radio galaxies \citep{Mag07}.
  Figure \ref{FS2images} shows image stamps of radio SN\,Ia host environments;
  galaxies within $\rm \Delta z \leq 0.1$ (i.e. $\rm \sim3\sigma_{\Delta z / (1+z)}$) are marked.
  This figure demonstrates that most radio loud SN\,Ia hosts have nearby neighbours, as expected for elliptical
  galaxies.
  As reported in Graham et al. (2008), the host of 05D1by is located in the outskirts of a galaxy cluster.
  We also find the galaxy density within $\rm r<50 kpc$ of this host shows clustering over the
  background field galaxy distribution, as determined by a ``significance parameter'' \citep{MG08};
  the small-scale environments of 04D1jg and 06D2ff hosts also show clustering, but not more so
  than the average SN\,Ia environment. In general we find the properties of radio SN\,Ia hosts
  and their environments are consistent with dusty starburst galaxies in non-isolated regions.
  A full analysis of SN\,Ia in small groups and pairs using spectroscopic
  (i.e. higher accuracy) galaxy redshifts will be presented in an upcoming paper.

  \section{CONCLUSION}
  \label{SConc}
  
  Based on the SNLS catalog, SN\,Ia rates in radio and infrared early-type galaxies
  are $\sim$1--5 times the rate in all early-type galaxies, but any enhancement is
  $\lesssim 2\sigma$. Rates in these subsets are consistent with predictions of the
  two component ``A+B'' SN\,Ia rate model. The infrared properties of SN\,Ia radio-loud
  early-type host galaxies suggest the presence of dust obscured star formation, and
  we have for the first time incorporated $\rm SFR_{IR}$ in the ``A+B'' model. In the C99 catalog,
  radio-loudest SN\,Ia hosts are consistent with the ``A+B'' model only if some SNe\,Ia
  and/or host galaxies have been misclassified. In general we find the properties of
  SN\,Ia in radio and infrared galaxies support a {\it continuum} of SN\,Ia and host
  galaxies properties, from young progenitors in very active galaxies to old progenitors
  in passive galaxies, as did DV05. Also these SN\,Ia are fainter and redder than other
  SN\,Ia, consistent with a dusty environment, and one was spectroscopically peculiar.
  
  To continue the investigation of SN\,Ia rates in different stellar populations,
  observations in the CFHT Deep fields at near- to far-infrared and sub-millimeter
  would be very useful. Radio coverage of the remaining two CFHTLS deep fields could
  double the number of high-redshift SN\,Ia in radio-loud early-type hosts available
  for analysis, but requires $\gtrsim 50$ hours per field to achieve the depth of
  VLA-VIRMOS (D1) and $\gtrsim 250$ hours per field for VLA-COSMOS depth (D2).
  Future work including spectroscopic galaxy redshifts will allow an analysis of the
  small scale environments of SN\,Ia hosts, and SN\,Ia rates and properties in small
  groups and interacting pairs.

  \acknowledgments
  
  We gratefully acknowledge the CFHT Queued Service Observations team, all SNLS collaboration members,
  Olivier Ilbert and Henry McCracken for early access to and correspondence regarding the photometric
  redshift galaxy catalog, and Enrico Cappellaro for access to the C99 galaxy and supernova samples.
  MLG acknowledges Colin Borys, Dave Patton, and Stephane Arnouts for their advice.
  This work is based in part on observations obtained with MegaPrime/MegaCam,
  a joint project of CFHT and CEA/DAPNIA, at the
  CFHT which is operated by the National Research Council (NRC) of Canada,
  the Institut National des Science de l'Univers of the Centre National de la
  Recherche Scientifique (CNRS) of France, and the University of Hawaii.
  This work is also based in part of data products produced at the Canadian Astronomy Data Centre
  as part of the CFHT Legacy Survey, a collaborative project of NRC and CNRS.
  This work is also based in part on on observations obtained at the Gemini Observatory,
  which is operated by the Association of Universities for Research in Astronomy, Inc.,
  under a cooperative agreement with the NSF on behalf of the Gemini partnership:
  the National Science Foundation (United States), the Science and Technology
  Facilities Council (United Kingdom), the National Research Council (Canada),
  CONICYT (Chile), the Australian Research Council (Australia),
  Minist\'{e}rio da Ci\^{e}ncia e Tecnologia (Brazil) and Ministerio de
  Ciencia, Tecnolog\'{i}a  e Innovaci\'{o}n  Productiva (Argentina)
  Gemini identification numbers of the programs under
  which these observations were taken are: GS-2003B-Q-8, GN-2003B-Q-9, GS-2004A-Q-11,
  GN-2004A-Q-19, GS-2004B-Q-31, GN-2004B-Q-16, GS-2005A-Q-11, GN-2005A-Q-11, GS-2005B-Q-6,
  GN-2005B-Q-7, GN-2006A-Q-7 and GN-2006B-Q-10.
  This research has made use of the NASA/ IPAC Infrared Science Archive, which is operated
  by the Jet Propulsion Laboratory,
  California Institute of Technology, under contract with the National Aeronautics and Space Administration.
  This publication has also made use of data products from the Two Micron All Sky Survey,
  which is a joint project of the University of Massachusetts and the Infrared Processing
  and Analysis Center/California Institute of
  Technology, funded by the National Aeronautics and Space Administration and the
  National Science Foundation.
  This work has been supported by NSERC and the University of Victoria.
  MLG gratefully acknowledges the financial support of the Province of British Columbia
  through the Ministry of Advanced Education.
  MS acknowledges support from the Royal Society.
  S. Fabbro acknowledges support from Funda\c{c}\~ao para a Ci\^{e}ncia e
  Tecnologia, Portugal, under grant SFRH/BPD/31817/006 and project POCTI/CTE-AST/57664/2004.

{\it Facilities:} \facility{CFHT, VLA, Spitzer, IRAS, Gemini, VLT, Keck}.

\clearpage

\clearpage
\begin{table}
\begin{center}
\caption{Properties of SNLS SN\,Ia and their Host Galaxies.  \label{TS2RLESSN}}
\begin{tabular}{lccccccccccc}
\tableline
SN\,Ia  & SN\,Ia     & SN\,Ia   & SN\,Ia    & Host & Host   & Host  & Host & Host Radio & Host IR    & Host IR   \\
SNLS ID & $\rm z_{spec}$  & Color    & Stretch   & SED & $\rm M_V$  & Mass  & sSFR & Luminosity & Luminosity & sSFR      \\ 
        &            &          &            &    &        & [$\rm 10^{10}\ M_{\odot}$] & [$\rm 10^{-10}\ y^{-1}$] & [$\rm 10^{29}\ erg\ s^{-1}\ Hz^{-1}$] & [$\rm 10^{10}\ L_{\odot}$] &  [$\rm 10^{-10}\ y^{-1}$] \\ 
\tableline						 							 
04D1jg  & 0.584      & -0.092   & $1.023\pm0.023$ & E/S0 & -22.99 & 43.5 & 0.15 & 14.5 & 46.47    & 1.56      \\
05D1by  & 0.299      &  0.736   & $1.040\pm0.061$ & Sbc  & -21.17 & 5.43 & 0.82 & 2.85 & 8.972    & 2.66      \\
05D1hn  & 0.149      &  0.334   & $1.049\pm0.020$ & E/S0 & -22.08 & 20.2 & 0.08 & 2.82 & 12.95    & 1.09      \\
06D2ck  & 0.552      & -0.020   & $1.083\pm0.048$ & E/S0 & -22.50 & 32.1 & 0.03 & 5.30 & $\ldots$ & $\ldots$  \\
06D2ff  & 0.345      &  0.299   & $0.980\pm0.092$ & Sbc  & -21.94 & 9.19 & 1.17 & 4.16 & 7.871    & 1.84      \\
06D2je  & 0.418      &  $\ldots$ & $\ldots$       & E/S0 & -19.79 & 2.16 & 0.25 & 1.73 & $\ldots$ & $\ldots$  \\
\hline                                                                                                     
04D2bt  & 0.220      &  0.222   & $0.988\pm0.062$ & E/S0 & -21.40 & 10.1 & 0.08 & 0.00 & 2.193    & 0.35      \\
06D2ca  & 0.531      &  0.047   & $1.086\pm0.049$ & Scd  & -21.03 & 0.81 & 6.83 & 0.00 & 33.56    & 24.6      \\
\tableline
\end{tabular}
\end{center}
\end{table}

\begin{table}
\begin{center}
\caption{Properties of C99 Elliptical Host Galaxies. \label{TDVE}}
\begin{tabular}{llcclccc}
\tableline\tableline
SN Ia Name                   & Galaxy                & Hubble                &  Radio                           & NED                          & Optical & Infrared   & Infrared \\
(and Type)\tablenotemark{a}  & Name\tablenotemark{a} & Type\tablenotemark{a} &  Luminosity\tablenotemark{a}     & Galaxy Type\tablenotemark{b} & sSFR    & Luminosity & sSFR     \\
                         &                       &                       &  [$\rm 10^{29}\ erg\ s^{-1}\ Hz^{-1}$] &                              & [$\rm 10^{-10}\ y^{-1}$] & [$\rm 10^{10}\ L_{\odot}$] & [$\rm y^{-1}$] \\
\tableline                      
1961H (Ia)  & NGC 4564  & -4.9  &  $<$0.008  ($<$0.010) & E6              & 0.00  &   $\ldots$  & $\ldots$ \\ 
1968A (I)   & NGC 1275  & -1.6  &  720.0     (1462)     & cD; pec; NLRG   & 0.26  &  17.5       & 0.63     \\ 
1970J (Ia)  & NGC 7619  & -5.0  &  0.540     (0.590)    & E               & 0.00  &   $\ldots$  & $\ldots$ \\ 
1972J (I)   & NGC 7634  & -2.0  &  $<$0.040  ($<$0.050) & SB0             & 0.27  &   $\ldots$  & $\ldots$ \\ 
1980I (Ia)  & NGC 4374  & -4.7  &  20.00     (31.18)    & E1; LINER; Sy2  & 0.00  &   0.2       & 0.02     \\ 
1980N (Ia)  & NGC 1316  & -1.9  &  1000.     (1.752)    & SAB; LINER      & 0.25  &   0.7       & 0.03     \\ 
1981D (Ia)  & NGC 1316  & -1.9  &  1000.     (1.752)    & SAB; LINER      & 0.25  &   0.7       & 0.03     \\ 
1981G (Ia)  & NGC 4874  & -4.0  &  24.00     (27.98)    & cD              & 0.03  &   $\ldots$  & $\ldots$ \\ 
1983G (Ia)  & NGC 4753  & -2.2  &  0.530     ($<$0.015) & I0              & 0.26  &   0.6       & 0.08     \\ 
1983J (?)   & NGC 3106  & -1.9  &  0.920     (0.692)    & S0              & 0.26  &   1.7       & 0.13     \\ 
1986G (Ia)  & NGC 5128  & -2.2  &  18.00     (52.39)    & S0; pec; Sy2    & 0.25  &   1.6       & 0.20     \\ 
1990M (Ia)  & NGC 5493  & -2.1  &  $<$0.040  ($<$0.051) & S0; pec         & 0.25  &   $\ldots$  & $\ldots$ \\ 
1991Q (?)   & NGC 4926A & -1.6  &  0.310     (0.325)    & S0; pec?; Sbrst & 0.24  &   6.2       & 2.86     \\ 
1991bg (Ia) & NGC 4374  & -4.7  &  20.00     (31.18)    & E1; LINER; Sy2  & 0.00  &   0.2       & 0.02     \\ 
1991bi (Ia) & NGC 5127  & -4.9  &  8.700     (9.843)    & E; pec          & 0.00  &   $\ldots$  & $\ldots$ \\ 
1992A (Ia)  & NGC 1380  & -1.9  &  $<$0.020  ($<$0.018) & SA0             & 0.25  &   0.4       & 0.06     \\ 
1992bo (Ia) & E352-G57  & -1.5  &  $<$0.160  ($<$0.175) & SB; pec         & 0.27  &   $\ldots$  & $\ldots$ \\ 
1993C (Ia)  & NGC 2954  & -4.9  &  $<$0.080  ($<$0.105) & E               & 0.00  &   $\ldots$  & $\ldots$ \\ 
1993ah (Ia) & E471-G27  & -2.0  &  41.00     (44.77)    & SBb             & 0.26  &   $\ldots$  & $\ldots$ \\ 
1994D (Ia)  & NGC 4526  & -2.0  &  0.042     (0.047)    & SAB             & 0.26  &   0.7       & 0.15     \\ 
1996X (Ia)  & NGC 5061  & -5.0  &  $<$0.030  ($<$0.033) & E0              & 0.00  &   $\ldots$  & $\ldots$ \\ 
\tableline
\tablenotetext{a}{From DV05, except radio powers in brackets from this paper.}
\tablenotetext{b}{Galaxy types (column 5) from the NASA Extragalactic Database (NED) where
  E: elliptical; cD: supergiant in cluster; I: irregular; LINER: low-ionization nuclear emission-line region;
  NLRG: narrow line radio galaxy; pec: peculiar; S: lenticular, A (unbarred), B (barred); Sbrst: starburst; Sy: Seyfert.}
\end{tabular}
\end{center}
\end{table}

\begin{table}
  \begin{center}
    \caption{Number of SN\,Ia in Early-Type Galaxies\tablenotemark{a} \label{TNSN}}
    \begin{tabular}{lcc}
      \tableline
      Sample   & SNLS\tablenotemark{b}      & C99 \\
      \tableline
      All           & $20\ (14.0^{+4.8}_{-3.7}) $  & $21^{+5.7}_{-4.5}$  \\
      Radio-loud    & $4 \ (2.8^{+2.9}_{-1.6})  $  & $9.5^{+4.2}_{-3.0}$ \\
      Radio-loudest & $1 \ (0.6^{+2.1}_{-0.6})  $  & $8^{+3.9}_{-2.8}$   \\
      BIRG          & $3 \ (2.0^{+2.6}_{-1.3})  $  & $4^{+3.2}_{-1.9}$   \\
      LIRG          & $2 \ (1.3^{+2.4}_{-1.0})  $  & $1^{+2.3}_{-0.8}$   \\
      \tableline
      \tablenotetext{a}{Uncertainties are derived from Poisson errors at the $1\sigma$ confidence level, as in Gehrels (1986).}
      \tablenotetext{b}{Brackets contain number of SN\,Ia corrected for SNLS detection efficiencies to be the number which exploded per year of SNLS observations, as in \S~\ref{sS2add}.}
    \end{tabular}
  \end{center}
\end{table}

\begin{table}
  \begin{center}
    \caption{Mass in Early-Type Galaxies\tablenotemark{a} \label{TMgal}}
    \begin{tabular}{lcc}
      \tableline
      Sample   & SNLS                    & C99 \\
      \tableline
      All           & 38440.4 (4337) & 23315.0 (2079) \\
      Radio-loud    & 3879.3  (165)  & 7724.5  (314)  \\
      Radio-loudest & 1638.8  (46)   & 2825.5  (92)   \\
      BIRG          & 1650.3  (179)  & 1805.7  (166)  \\
      LIRG          & 747.0   (73)   & 469.2   (22)   \\
      \tableline
      \tablenotetext{a}{Brackets contain number of galaxies in sample.}
    \end{tabular}
  \end{center}
\end{table} 

\begin{table}
  \begin{center}
    \caption{SN\,Ia Rates in Early-Type Galaxies\tablenotemark{a} \label{TSNRates}}
    \begin{tabular}{lcc}
      \tableline
      Sample   & SNLS                    & C99 \\
               & [SNuM\tablenotemark{b}]                  & [SNuM\tablenotemark{b}] \\
      \tableline
      All           & $0.051^{+0.018}_{-0.014}$ & $0.044^{+0.012}_{-0.010}$ \\
      Radio-loud    & $0.102^{+0.103}_{-0.057}$ & $0.101^{+0.045}_{-0.032}$ \\
      Radio-loudest & $0.055^{+0.186}_{-0.051}$ & $0.197^{+0.097}_{-0.068}$ \\
      BIRG          & $0.168^{+0.217}_{-0.107}$ & $0.166^{+0.131}_{-0.079}$ \\
      LIRG          & $0.249^{+0.469}_{-0.191}$ & $0.213^{+0.488}_{-0.177}$ \\
      \tableline
      \tablenotetext{a}{Uncertainties are derived from Poisson errors at the $1\sigma$ confidence level, as in Gehrels (1986).}    
      \tablenotetext{b}{[SNuM] = $\rm SN\ (100\ yr)^{-1}\ (10^{10}\ M_{\odot})^{-1}$}
    \end{tabular}
  \end{center}
\end{table}

\begin{table}
  \begin{center}
    \caption{Ratios of SN\,Ia Rates in Early-Type Galaxies\tablenotemark{a} \label{TSNRatios}}
    \begin{tabular}{lcc}
      \tableline
      Sample   & SNLS              & C99 \\
      \tableline
      Radio-loud    & $2.0^{+3.5}_{-1.3} $  & $2.3^{+1.9}_{-1.1}$ \\
      Radio-loudest & $1.1^{+5.3}_{-1.0} $  & $4.5^{+4.0}_{-2.1}$ \\
      BIRG          & $3.3^{+6.9}_{-2.4} $  & $3.7^{+4.8}_{-2.2}$ \\
      LIRG          & $4.9^{+14.2}_{-4.0}$  & $4.8^{+15.3}_{-4.2}$ \\
      \tableline
      \tablenotetext{a}{Uncertainties are derived from Poisson errors at the $1\sigma$ confidence level, as in Gehrels (1986).} 
    \end{tabular}
  \end{center}
\end{table} 

\begin{table}
  \begin{center}
    \caption{Statistical Comparison to ``A+B'' SN\,Ia Rate Model\tablenotemark{a} \label{TPP} }
    \begin{tabular}{llccccc}
      \tableline
      \multicolumn{2}{c}{Sample} & $\rm N_{obs}$ & $\rm N_{A+B,opt}$ & $\rm P_{opt}$ & $\rm N_{A+B,IR}$ & $\rm P_{IR}$ \\
      \tableline
      Radio-loud    & SNLS  & 4   & 2.16  & 0.17   & 2.38  & 0.22  \\
                    & C99   & 9.5 & 4.84  & 0.04   & 5.10  & 0.05  \\
      \tableline
      Radio-loudest & SNLS  & 1   & 0.90  & 0.59   & 1.00  & 0.73  \\
                    & C99   & 8   & 2.35  & 0.003\tablenotemark{b}  & 2.39  & 0.003\tablenotemark{b} \\
      \tableline
      BIRG          & SNLS  & 3   & 1.03  & 0.09   & 1.52  & 0.20  \\
                    & C99   & 4   & 1.39  & 0.05   & 1.77  & 0.10  \\
      \tableline
      LIRG          & SNLS  & 2   & 0.44  & 0.07   & 0.83  & 0.20  \\
                    & C99   & 1.0 & 0.26  & 0.23   & 0.43  & 0.35  \\
     \tableline   
     \tablenotetext{a}{Columns defined in \S~\ref{SABP}.}
     \tablenotetext{b}{Further considerations raise $\rm P\gtrsim0.1$ in \S~\ref{sSABPradio}.}
    \end{tabular}
  \end{center}
\end{table}

\end{document}